\documentclass[12pt]{article}
\pdfoutput=1                     
\usepackage{cite}
\usepackage[usenames]{color}
\usepackage{graphicx,amssymb,bbold,bm}
\usepackage{amsmath,amsfonts,epsfig}
\usepackage{braket}
\usepackage[usenames]{color} 
\usepackage{wrapfig}
\usepackage{amsfonts,bm}
\usepackage{epsfig,amsfonts,bm}
\usepackage{tikz}
\usepackage{ulem}
\usepackage{here}

\newcommand{\pslash}{p\kern-1ex /}
\newcommand{\qslash}{q\kern-1ex /}
\newcommand{\lslash}{l\kern-1ex /}
\newcommand{\sslash}{s\kern-1ex /}
\newcommand{\kaslash}{k_a\kern-2ex /}
\newcommand{\kbslash}{k_b\kern-2ex /}
\newcommand{\Dslash}{{\cal D}\kern-1.5ex /}

\newcommand{\bc}{\overline{c}}

\newcommand{\beqa}{\begin{eqnarray}}
\newcommand{\eeqa}{\end{eqnarray}}

\renewcommand{\Im}{\mathrm{Im}\,}
\newcommand{\bpm}{\begin{pmatrix}}
\newcommand{\epm}{\end{pmatrix}}
\newcommand{\bbm}{\begin{bmatrix}}
\newcommand{\ebm}{\end{bmatrix}}

\newcommand{\f}{\frac}
\newcommand{\Real}{\mathbb{R}}
\newcommand{\AdSt}{{\text{AdS}$_3$\,\,}}
\makeatletter

\@addtoreset{equation}{section}
\makeatother
\usepackage{subfig}
\begin{document}

\voffset -0.7 true cm
\hoffset 1.5 true cm
\topmargin 0.0in
\evensidemargin 0.0in
\oddsidemargin 0.0in
\textheight 8.6in
\textwidth 5.4in
\parskip 9 pt
 
\def\Tr{\hbox{Tr}}
\newcommand{\be}{\begin{equation}}
\newcommand{\ee}{\end{equation}}
\newcommand{\bea}{\begin{eqnarray}}
\newcommand{\eea}{\end{eqnarray}}
\newcommand{\beas}{\begin{eqnarray*}}
\newcommand{\eeas}{\end{eqnarray*}}
\newcommand{\nn}{\nonumber}
\font\cmsss=cmss8
\def\C{{\hbox{\cmsss C}}}
\font\cmss=cmss10
\def\bigC{{\hbox{\cmss C}}}
\def\scriptlap{{\kern1pt\vbox{\hrule height 0.8pt\hbox{\vrule width 0.8pt
  \hskip2pt\vbox{\vskip 4pt}\hskip 2pt\vrule width 0.4pt}\hrule height 0.4pt}
  \kern1pt}}
\def\ba{{\bar{a}}}
\def\bb{{\bar{b}}}
\def\bc{{\bar{c}}}
\def\bphi{{\Phi}}
\def\Bigggl{\mathopen\Biggg}
\def\Bigggr{\mathclose\Biggg}
\def\Biggg#1{{\hbox{$\left#1\vbox to 25pt{}\right.\n@space$}}}
\def\n@space{\nulldelimiterspace=0pt \m@th}
\def\m@th{\mathsurround = 0pt}

\begin{titlepage}
\begin{flushright}
{\small OU-HET-1081, YITP-20-158} 
 \\
\end{flushright}

\begin{center}

\vspace{5mm}

{\Large \bf Wormholes and holographic decoherence} \\[3pt] 
\vspace{1mm}

\vspace{7mm}

\renewcommand\thefootnote{\mbox{$\fnsymbol{footnote}$}}
Takanori Anegawa\footnote{takanegawa@gmail.com}$^{\spadesuit}$, 
Norihiro Iizuka\footnote{iizuka@phys.sci.osaka-u.ac.jp}$^{\spadesuit}$,   
\\
\vspace{1mm}
Kotaro Tamaoka\footnote{kotaro.tamaoka@yukawa.kyoto-u.ac.jp}$^{\diamondsuit}$  
and 
Tomonori Ugajin\footnote{tomonori.ugajin@yukawa.kyoto-u.ac.jp}$^{\diamondsuit,\clubsuit}$
\vspace{3mm}

${}^\spadesuit$
{\small \sl Department of Physics, Osaka University} \\ 
{\small \sl Toyonaka, Osaka 560-0043, JAPAN}

${}^\diamondsuit$
{\small \sl Center for Gravitational Physics, Yukawa Institute for Theoretical Physics,} \\ 
{\small \sl  Kyoto University, Kyoto 606-8502, JAPAN} 

${}^\clubsuit$
{\small \sl The Hakubi Center for Advanced Research, Kyoto University,
} \\ 
{\small \sl Kyoto 606-8501, JAPAN} 

\end{center}

\noindent

\vspace{-7mm}

\abstract{We study a class of decoherence process which admits a 3 dimensional holographic bulk. 
Starting from a thermo-field double dual to a wormhole, we prepare another thermo-field double which plays the role of environment. By allowing the energy flow between the original and environment thermo-field double, the entanglement of the original thermo-field double eventually decoheres. We model this decoherence by four-boundary wormhole geometries, and study the time-evolution of the moduli parameters to see the change of the entanglement pattern among subsystems.  A notable feature of this holographic decoherence processes is that at the end point of the processes, the correlations of the original thermo-field double are lost completely both classically and also quantum mechanically.  We also discuss distinguishability between thermo-field double state and thermo mixed double state, which contains only classical correlations, and construct a code subspace toy model for that.} 
\\

\end{titlepage}

\setcounter{footnote}{0}
\renewcommand\thefootnote{\mbox{\arabic{footnote}}}

\newpage

\setcounter{tocdepth}{2}  

\section{Introduction and summary}

Wormholes are quite interesting geometries which connect  disjoint spacetimes.  Even though their phenomenological role has been discussed long \cite{Coleman:1988cy,Hawking:1987mz,Hawking:1988ae,Giddings:1987cg,Giddings:1988cx,Giddings:1988wv,Lavrelashvili:1987jg} (See also \cite{Hebecker:2018ofv} for a review), recently wormholes was paid attention again since they refined our understanding of holography and quantum gravity.  
To see the roles of Euclidian wormholes in holography, 
let us consider boundary theory partition function where the boundary has several disjoint components and assuming no interactions between disjoint components.  Then the boundary partition function should be factorized into the contributions of  each boundary component. On the other hand, the gravitational path integral is not,  if we take into account the contributions of the Euclidean wormholes connecting disjoint boundaries through the emergent bulk. 
This is the factorization paradox pointed out in \cite{Maldacena:2004rf}, where such Euclidian wormholes connecting multiple disjoint boundaries are explicitly constructed.

\vspace{0.2cm}

It was found that this non-factorization is naturally understood, once we interpret the boundary theory is not a single theory,  but rather an ensemble of boundary theories. Indeed,  in a  remarkable paper \cite{Saad:2019lba}, it was shown that  JT gravity path integral shows the ensemble nature of the boundary theory, once we include all possible wormholes dual to a matrix model.    Also \cite{Marolf:2020xie} studied a toy model of  gravitational path integral,  providing a strong  evidence of the claim. See also \cite{ArkaniHamed:2007js, Balasubramanian:2020jhl,Gardiner:2020vjp,Gesteau:2020wrk,Belin:2020hea,Maloney:2020nni,Afkhami-Jeddi:2020ezh,Cotler:2020ugk,Altland:2020ccq}
 for related discussions.

\vspace{0.2cm}

In addition to these effects of Euclidean wormholes,  Lorentzian wormholes also play an important role in quantum gravity.   
One of the concrete examples of such Lorentzian wormhole is the Einstein Rosen (ER) bridge of an eternal black hole, which connects two asymptotic regions. 
Motivated by the holographic correspondence between this ER bridge and thermo-field double state in the dual field theory \cite{Maldacena:2001kr,Israel:1976ur,Balasubramanian:1998de}, 
it was conjectured in such ER bridge is a geometric manifestation of entanglement structure of the  underlying state in quantum gravity\cite{Maldacena:2013xja}.   This conjecture, called ER= EPR, has been applied to various topics on the relation between quantum information theory and gravity.

\vspace{0.2cm}

Most remarkable point of this ER=EPR conjecture is that it  suggests  that quantum entanglement is an indispensable ingredient for the emergence of smooth  geometry in the  semi-classical limit of gravity \cite{VanRaamsdonk:2010pw}.  On the other hand, correlation is not always induced quantum mechanically, and it might be possible that classical correlation can induce similar effects.   
Therefore to understand the conjecture more, a natural question we would like to address in this paper is;  {\it instead of quantum entanglement, can classical correlation have such a smooth geometric description in dual gravity?}   
Recently Verlinde in \cite{Verlinde:2020upt} argued that classical correlation can also have an ER bridge type of smooth geometry.

 \vspace{0.2cm}

\begin{figure}[htbp]
 \begin{center}
  \resizebox{120mm}{!}{\includegraphics{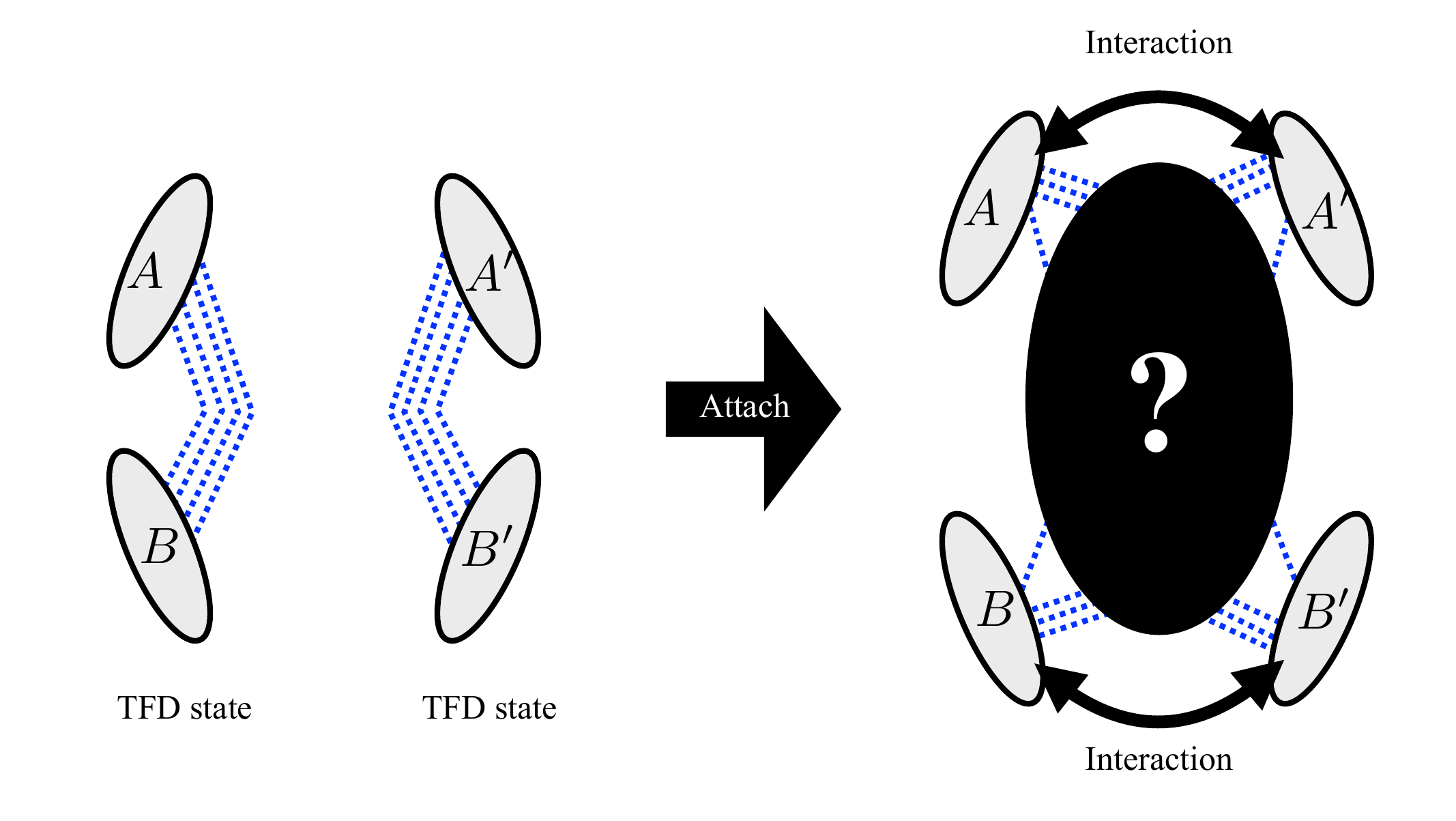}}\\\resizebox{110mm}{!}{\includegraphics{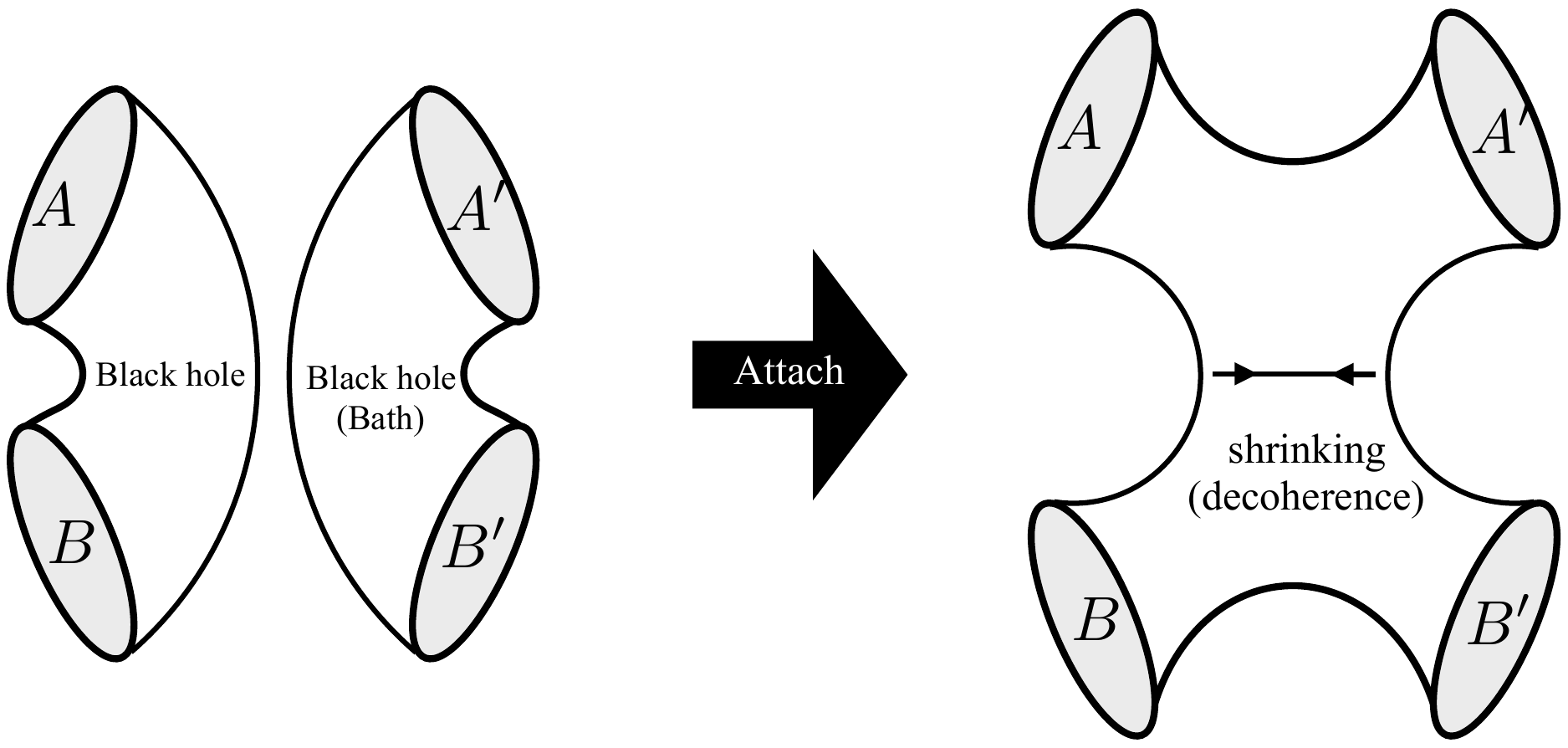}}
 \end{center}
 \caption{Our toy model of holographic decoherence. We start from two thermo-field double states (upper-left panel) and its dual two-sided blackholes (lower-left panel). Here blue dotted lines in the upper-left panel represent the entanglement. First we attach these systems by introducing an interaction between $A$ and $A^\prime$, $B$ and $B^\prime$ as well. Then, these interactions cause an energy flow between $A$ and $A^\prime$, and one between $B$ and $B^\prime$. We would like to ask the entanglement structure in the final process of this decoherence (upper-right panel). Our process can be holographically viewed as a shrinking of a neck between $A$ and $B$ (or $A^\prime$ and $B^\prime$) which represents a magnitude of the entanglement between them (lower-right panel).}
 \label{fig:setup}
\end{figure}

To see if an ER bridge can have only classical correlation, 
we consider the following decoherence process. 
Let us start from an AdS eternal black hole. The ER bridge of the eternal black hole is induced purely by quantum entanglement, 
since this two-sided  eternal black hole is dual to a  thermo-field double state on a bipartite system. Let us call this bi-partite system as $A$ and $B$.  See Figure \ref{fig:setup}.    
We then prepare an auxiliary bipartite system $A'$ and $B'$  which is again modeled  by another eternal black hole. This auxiliary bipartite system $A'$ and $B'$ plays the role of heat bathes/environment. We then attach this auxiliary black hole ($A'$ and $B'$) to the  original two sided  black hole ($A$ and $B$) and allow the energy flow from $A$ to $A'$, and similarly, from $B$ to $B'$.  
In the dual conformal field theory point of view, this process induces equilibration
between $A$ and $A'$ and similarly $B$ and $B'$ and simultaneously, 
induces decoherence between $A$ and $B$.
What we would like to see is, as the initial thermo-field double state ($A$ and $B$) interacts with heat bathes ($A'$ and $B'$), how the original quantum entanglement between $A$ and $B$ can be washed out, and leave, even if exist, only classical correlation. 

In this paper, we concretely study this  decoherence process in the $AdS_{3}$/CFT$_{2} $ setup. It is well known that  in  $AdS_{3}$ gravity contains various multi-boundary wormholes which are natural generalizations of two sided BTZ black hole  to the setups with multiple boundaries.  Moreover, thanks to the topological nature of pure  $AdS_{3}$ gravity, this class of  multi-boundary wormholes are constructed by taking quotient of the pure $AdS_3$ \cite{Brill:1998pr,Skenderis:2009ju,Aminneborg:1997pz,Balasubramanian:2014hda}.  Therefore we model the decoherence  process described in the previous paragraph, by such a wormhole with four boundaries.   
This four boundary wormhole is dual to a pure state on $ABA'B'$ with interesting entanglement structure \cite{Balasubramanian:2014hda}. Other previous studies on entanglement in the class of multi-boundary wormholes include
\cite{Maxfield:2014kra,Zhang:2016evx,Balasubramanian:2020hfs,Li:2020ceg, Dai:2020ffw}.
By tracing out the thermal bath $A'B'$, we in general obtain  a mixed state on $AB$.  
Actual time evolution is  modeled by changing the masses of four horizons  in the geometry, whose total is kept fixed.  
Under the time evolution, we expect the the quantum entanglement originally shared between $A$ and $B$ is eventually  swapped to the entanglement between $A$ and $A'$  and that of between $B$ and $B'$.  
The monogamy of entanglement suggests, the correlation in the final state, if it is still left, is only classical.  

\vspace{0.2cm}

However, we  find that the final state of the holographic decoherence process can not have{ \it any } correlation between $A$ and $B$, 
both classically and quantum mechanically.  
This in particular means that we cannot construct an ER bridge which only contains classical correlation.  
Given that CFTs need to be sufficiently chaotic \cite{Shenker:2014cwa,Maldacena:2015waa} to have holographic bulk dual, we believe that this loss of correlation holds generically in holographic system. 
Indeed in such a theory, we expect in the final state, the mutual information between the original black hole $A$ and the bath $A'$ gets maximized. 
Then the monotonicity of mutual information tells us that the mutual information of original thermo-field double, namely between $A$ and $B$, has to vanish, implying there is no correlation between $A$ and $B$ at the final stage of the process.

\vspace{0.2cm}

This observation does not exclude the possibility that the classical correlation is indistinguishable from quantum entanglement, when the bulk observer can only probe code subspace, which is low energy subspace of  the total  quantum gravity Hilbert space.  In the latter half of this paper, we explore this possibility in the toy model.

\vspace{0.2cm}

This paper is organized as follows. In section \ref{sec:wh}, we review how to construct multi-boundary wormholes and in section \ref{sec:decohe}, we study time evolution of the decoherence process through the moduli change of the four-boundary wormholes and show that after holographic decoherence, any correlation, both classical and quantum one, are left. Section \ref{sec:toy} we discuss distinguishability of thermo mixed double state with thermo-field double state, and also toy model for that.
Section \ref{sec:disc} is discussion. In the appendix \ref{AppA}, we summarize the known properties of $AdS_3$ and constructions of multi-boundary wormholes there.

\section{Construction of wormholes by quotients} \label{sec:wh}

We would like to discuss a model of holographic decoherence by using four wormhole geometries in $AdS_3$. 
Imagine starting from two two-sided black holes (see left panel of Figure \ref{fig:setup}). One two-sided black hole has two boundaries $A$ and $B$, whereas another one does $A^\prime$ and $B^\prime$. We will treat the boundaries of the latter two-sided black hole as two thermal bathes. We then attach each heat bath ($A^\prime$ and $B^\prime$) to one of the boundaries ($A$ and $B$) of the former black hole, and allow energy flows between them. Then the initial entanglement shared between $A$ and $B$ starts to spread over the entire four party system. In particular, the entanglement between $A$ and $A^\prime$ ($B$ and $B^\prime$) would become large, whereas the original entanglement between $A$ and $B$ ($A^\prime$ and $B^\prime$) would get lost (see right panel of Figure \ref{fig:setup}). Hence, the resulting reduced density matrix $\rho_{AB}$ is expected to reach a separable state, where only classical correlations are left. In a forthcoming section, we would like to check this statement explicitly by computing holographic entanglement entropies in a specific four boundary wormhole system.  

In order to construct such multi-boundary wormholes, it is useful to start from the hyperbolic slicing of it, with the metric, 
\be 
ds^{2} =-dt^{2} +\cos^{2} t \; d^{2}s_{\mathbb{H}^{2}}. \label{eq:metric}
\ee
We can construct $t=0$ time slice of a wormhole from a quotient of two-dimensional hyperbolic space $\mathbb{H}^{2}/\Gamma$. Here $\Gamma$ is a discrete subgroup of its isometry, $SL(2,\mathbb{R})$\footnote{Since we focus on the time slice, we focus only on the isometry of $\mathbb{H}^{2}$. This $SL(2,\mathbb{R})$ is a diagonal subgroup of the original isometry for $AdS_{3}$. }. 
Such wormhole construction has been well-studied in literature, see for example,  \cite{Brill:1998pr,Skenderis:2009ju,Aminneborg:1997pz,Balasubramanian:2014hda}. For the sake of clarity and fixing our notation, here we will discuss two-, three- and four-boundary wormholes in order. The reader who is familiar with the construction of wormhole geometries can skip the rest part of this section. Unfamiliar one may also see the appendix \ref{AppA}. 

\subsection*{A) Eternal BTZ black holes} 

\begin{figure}[t]
 \hspace{-12mm}\resizebox{80mm}{!}{\includegraphics{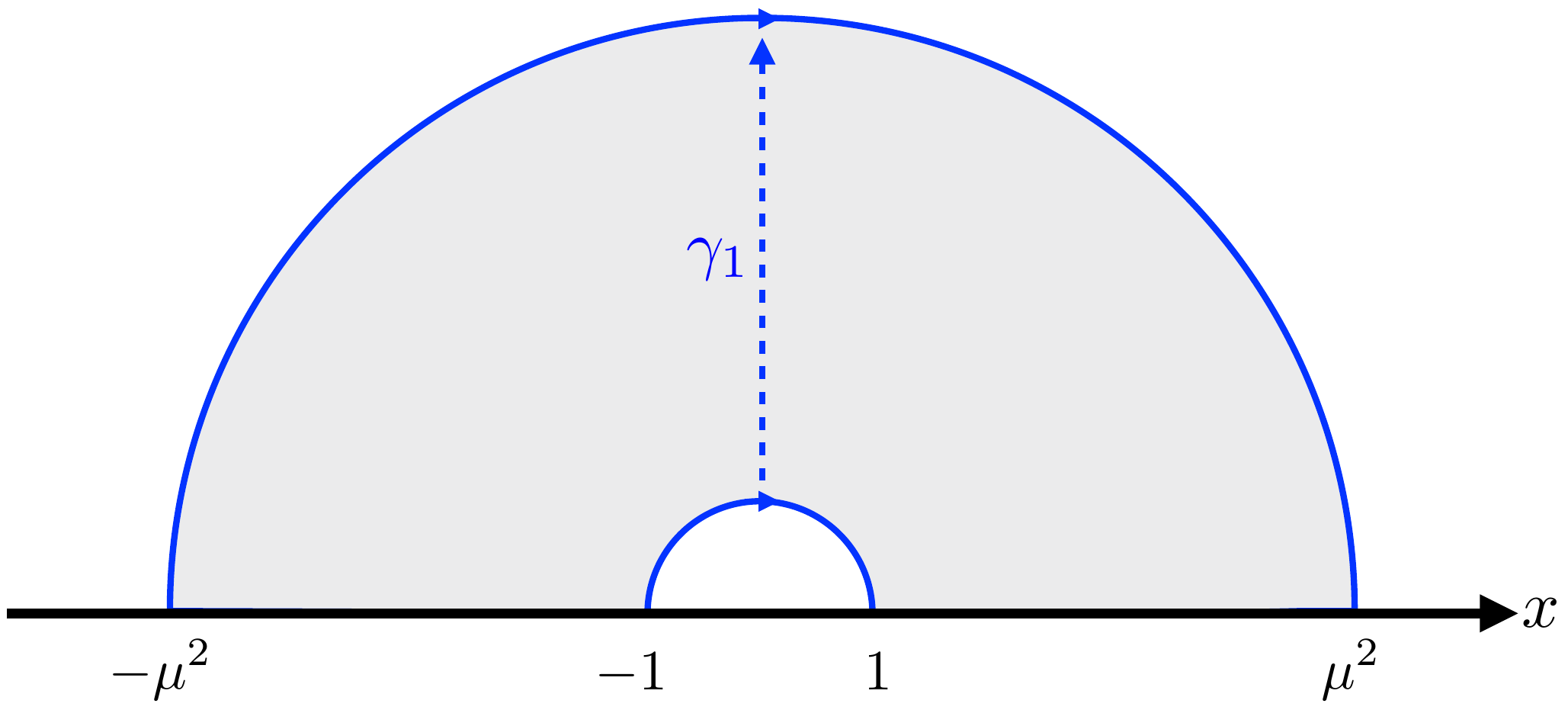}}\resizebox{80mm}{!}{\includegraphics{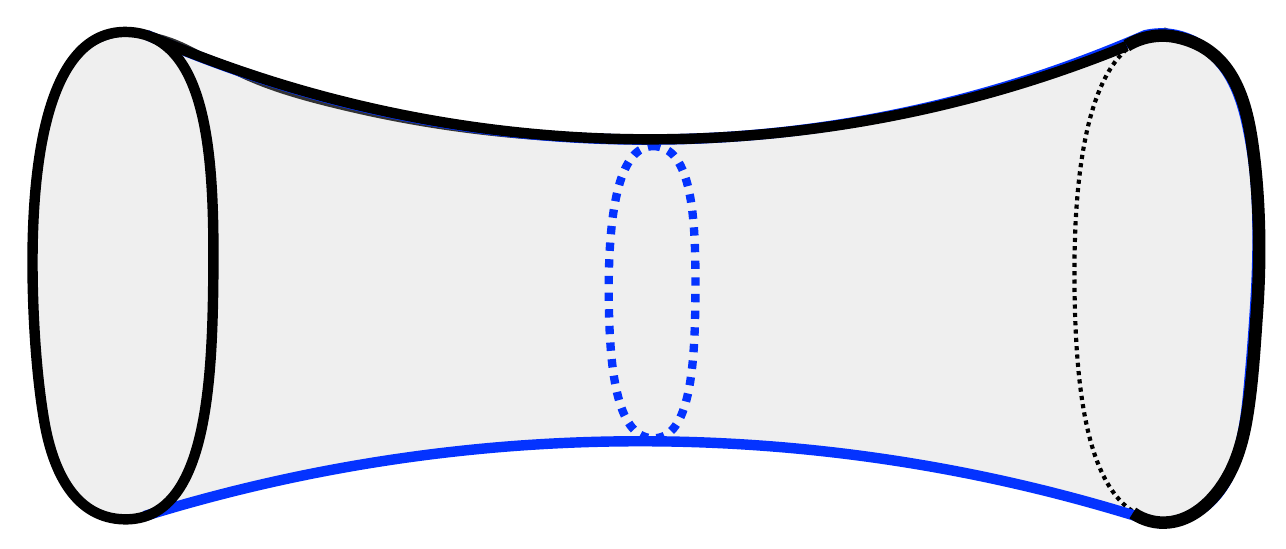}}
 \caption{Left: We display the upper half-plane and the Poincare coordinates. Gray shaded region is a fundamental domain of the identification by $\gamma_1$. In particular, the smaller blue semi-circle is identified by larger one. Right: After the identification, resulting cylindrical geometry becomes a canonical timeslice of the (static) eternal BTZ. The area of horizon (the blue dotted curve in center) can be computed as the length of blue dotted line in the left panel. }
  \label{fig:btz}
\end{figure}

The simplest example is the eternal BTZ black hole. In what follows, we will take the Poincare coordinates as a coordinate of $\mathbb{H}^2$, 
\be
d^{2}s_{\mathbb{H}^{2}}=\dfrac{dz^2+dx^2}{z^2}.
\ee
We will construct a time slice of a BTZ black hole by a quotients of this Poincare coordinates (namely, an upper half plane). 
The corresponding group $\Gamma$ only has one generator $\gamma_1$, which can be diagonalized 
\be 
\gamma_1 =\left(
    \begin{array}{cc}
      \mu & 0 \\
      0 &  \frac{1}{\mu} \\
    \end{array}
  \right),
\ee
which acts as a dilatation for Poincare coordinates $x \rightarrow \mu^2 x, z \rightarrow \mu^2z $. We shall take $\mu=e^{ \pi r_{+}}$ since as we will see, then the area of horizon becomes $2\pi r_+$.   
The fundamental region after the quotients can be chosen to be the region bounded by two semi-circles, 
\be
C_a: x^{2} +z^{2} =1, \;\;\;C_b: x^{2} +z^{2} =\mu^4,  
\ee
and the boundary of the upper half plane $z=0$. After these two semicircles are identified, we obtain a cylinder which is the time slice of the eternal BTZ black hole. See Figure \ref{fig:btz}.  

We can compute the area of the horizon $L_1\equiv L(\gamma_1)$ from the minimal distance between $C_a$ and $C_b$ (centered blue dotted line in the left panel of Figure \ref{fig:btz}). In our setup, it is easily computed as the length of the straight line,  
\be 
L_1 = \int^{\mu^2}_{1}  \f{dz}{z} =2\log \mu =2\pi r_+. \label{eq:line}
\ee
For more general minimal distance, one may also refer the formula in Appendix \ref{AppA}. From the viewpoint of group elements, this simplification happens thanks to the diagonalization of $\gamma_1$. 
In particular, one can also relate the $L_1$ to a trace of the group element $\gamma_1$,
\be 
\dfrac{1}{2}\mathrm{tr} \gamma_{1} =\cosh \f{L_1}{2}. \label{eq:Lbtz}
\ee
Applying an appropriate coordinate transformation, one can also relate the length of more general closed geodesics to a straight line computation as the present BTZ example. Therefore, we can also compute more general ones from the trace of a certain group element as \eqref{eq:Lbtz}. In what follows, we denote $L(\gamma)$ as a length which can be calculated as
\be
\dfrac{1}{2} | \mathrm{tr} \gamma | =\cosh \f{L(\gamma)}{2}. \label{eq:formula}
\ee

See the appendix \ref{App5} for the derivation of this eq.

\subsection*{B) Three boundary wormholes}
\begin{figure}[t]
 \hspace{-12mm}\resizebox{100mm}{!}{\includegraphics{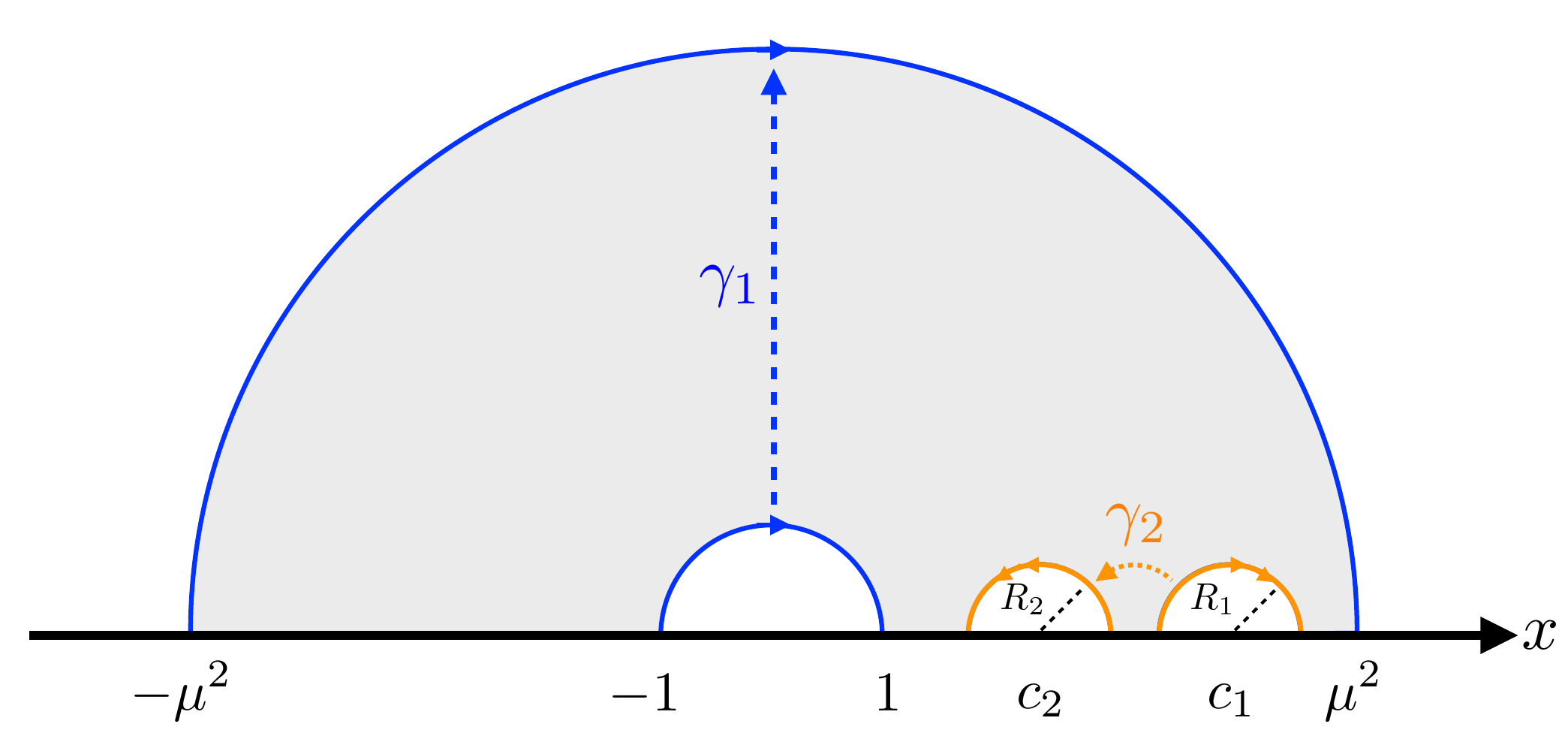}}\resizebox{60mm}{!}{\includegraphics{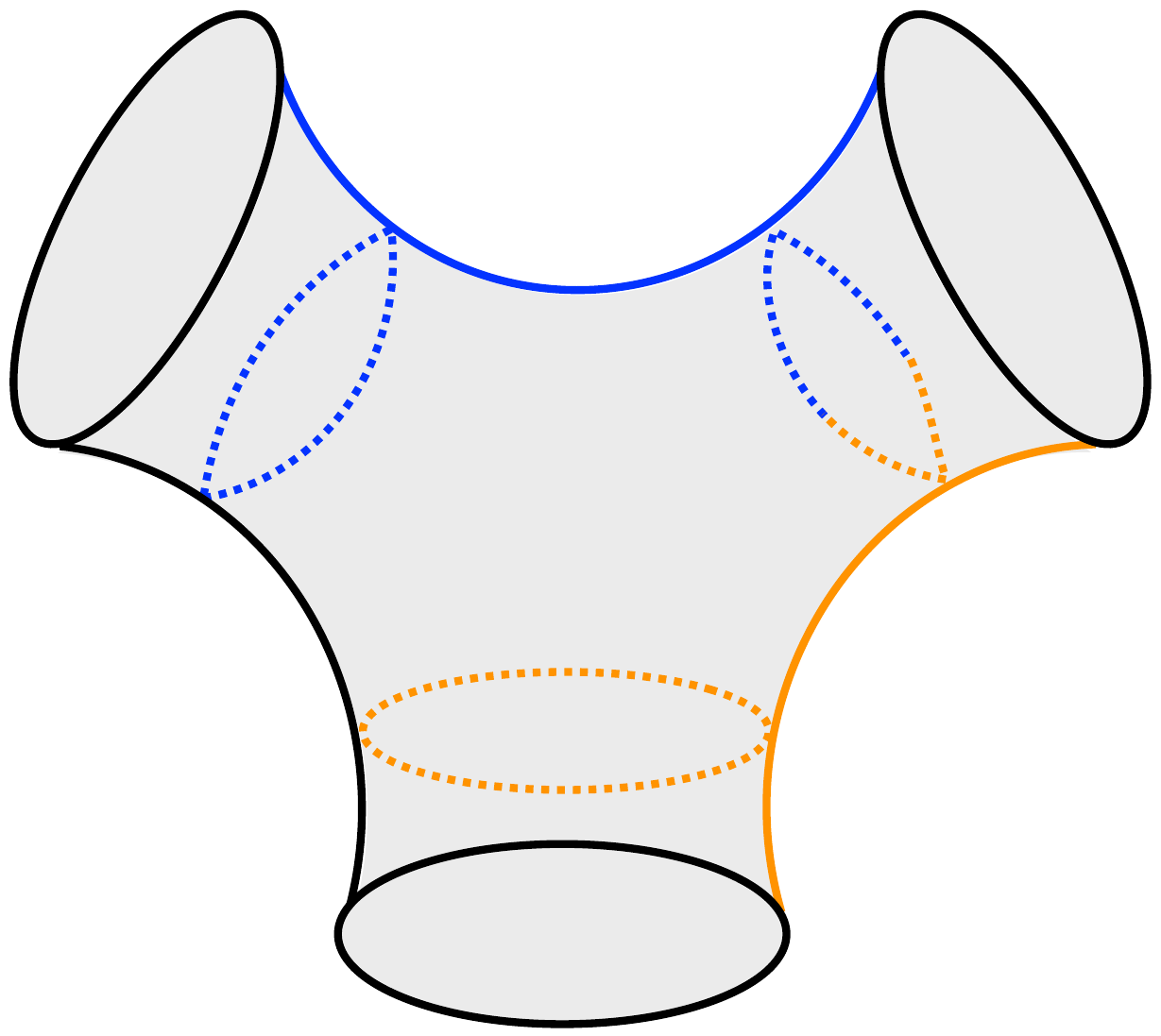}}
 \caption{Three-boundary wormhole obtained from the identification \eqref{eq:ids3}. Left: Gray shaded region is a fundamental domain of the identification by $\{\gamma_1,\gamma_2\}\in\Gamma$. Colored circles with arrows will be seen as identical to each other, including their orientation. Right: The resulting three boundary-wormhole. A blue (An orange) circle will be calculated from $\mathrm{tr} \gamma_{1}$ ($\mathrm{tr} \gamma_{2}$). A half blue and half orange circle will be obtained via ${\mathrm{tr} (\gamma_{1}\gamma_{2})}$. }
  \label{fig:3bdry}
\end{figure}

Next, let us consider a slightly involved example, a three-boundary wormhole. In doing so, we consider identification group $\Gamma$ generated by the previous $\gamma_{1}$ and
\be 
\gamma_{2}=\left(
    \begin{array}{cc}
     -\f{c_{2}}{\sqrt{R_{1}R_{2}}}& \f{c_1c_2+R_1R_2}{\sqrt{R_{1}R_{2}}} \\
     -\f{1}{\sqrt{R_{1}R_{2}}}  & \f{c_{1}}{\sqrt{R_{1}R_2}} \\
    \end{array}
  \right).
  \label{eq:ids3}
\ee
The role of $\gamma_{1}$ is just the same as previous example, namely it identifies $C_a$ and $C_b$. The second generator $\gamma_2$ consists of two translations, an inversion and a dilatation so that it can identify
\be
C_1: (x-c_{1})^{2} +z^{2} =R^{2}_{1}, \; C_2: (x-c_{2})^{2} +z^{2} =R^{2}_{2},
\ee
with the flip of orientation. For more detail, see around \eqref{eq:total} of the appendix \ref{AppA4}.  
Then, the fundamental domain of $\mathbb{H}^2/\Gamma$ is the region bounded by these four circles as well as the boundary of the upper half plane. See Figure \ref{fig:3bdry}. 

Then we have three horizons associated with each boundary. Following the previous formula \eqref{eq:formula}, these area can be calculated via $\mathrm{tr}\gamma_{1},\mathrm{tr} \gamma_{2}$, and ${\mathrm{tr} \gamma_{1}\gamma_{2}}$ respectively, 
\begin{align}
L(\gamma_1)&=2\log\mu, \\ 
L(\gamma_2)&=2\cosh^{-1}\left[\frac{ c_{1}-c_{2} }{2\sqrt{R_{1}R_{2}}}\right], \\ 
L(\gamma_{1}\gamma_{2})&=2\cosh^{-1}\left[\left|\frac{{c_{1}\mu^{-1}-c_{2}\mu}}{2\sqrt{R_{1}R_{2}}}\right|\right].
\end{align}

\subsection*{C) Four boundary wormholes}

\begin{figure}[t]
 \begin{center}
 \resizebox{140mm}{!}{\includegraphics{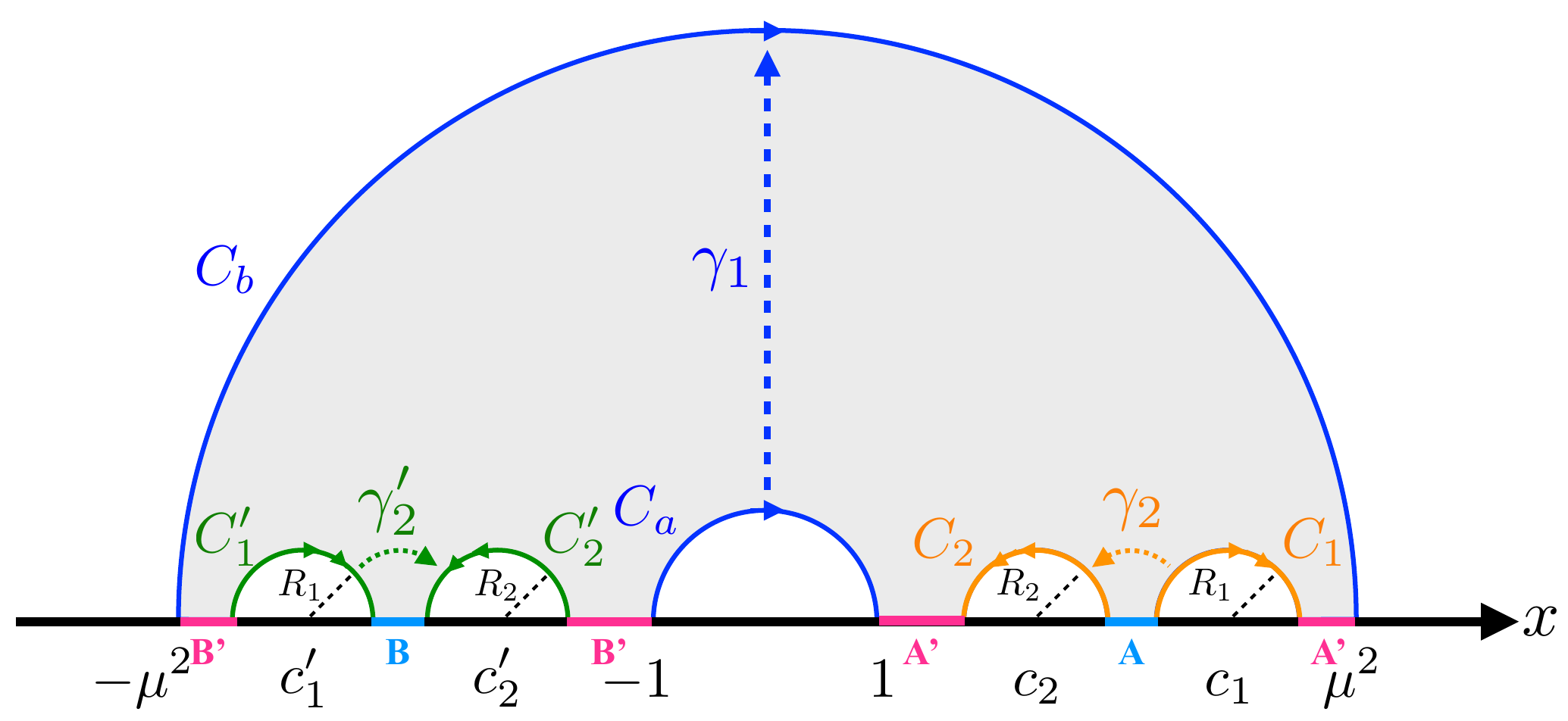}}
 \end{center}
 \caption{Four boundary wormhole of our interest. Here we assign the boundary subregions $A=[c_2+R_2,c_1-R_1], B=[c_1^\prime+R_1^\prime,c_2^\prime-R_2^\prime], A^\prime=[1,c_2-R_1]\cup[c_1+R_1,\mu^2], A^\prime=[-\mu^2,c^\prime_1-R_1^\prime]\cup[c_2^\prime+R_2,-1]$. The resulting four boundary-wormhole has a shape of lower-right panel of Figure \ref{fig:setup}.}
 \label{fig:4bdry}
\end{figure}

The extension to four boundary wormholes of our interest is almost straightforward. For the identification, we reuse $\gamma_1$ and $\gamma_2$ in the previous examples. We also include the following group element,
\be
\gamma_2^\prime=\bpm
-\frac{c^\prime_2}{\sqrt{R^\prime_1R^\prime_2}}& \frac{c^\prime_1c^\prime_2+R^\prime_1R^\prime_2}{\sqrt{R^\prime_1R^\prime_2}} \vspace{2mm}\\
     -\frac{1}{\sqrt{R^\prime_1R^\prime_2}}  & \frac{c^\prime_1}{\sqrt{R^\prime_1R^\prime_2}} \\
\epm,
\ee
where $c_i^\prime$ and $R^\prime_i$ corresponds to position of center and radius for semi-circles $C_1^\prime$ and $C_2^\prime$ (see Figure \ref{fig:4bdry}). 
Clearly $\gamma^\prime_2$ plays the same role as $\gamma_2$. Following the same prescription as the three boundary wormhole, it is straightforward to compute the area of horizons,
\begin{align}
L_{A} &\equiv L(\gamma_2)=2\cosh^{-1}\left[\frac{ c_{1}-c_{2} }{2\sqrt{R_{1}R_{2}}}\right], \\ 
L_{A^\prime}&\equiv {L(\gamma_{1}\gamma_{2})=2\cosh^{-1}\left[\left|\frac{ c_{1}\mu^{-1}-c_{2}\mu }{2\sqrt{R_{1}R_{2}}}\right|\right]}, \\
L_B &\equiv L({\gamma^\prime_2})= 2\cosh^{-1}\left[\frac{ -c^\prime_{1}+c^\prime_{2} }{2\sqrt{R^\prime_{1}R^\prime_{2}}}\right], \\
L_{B^\prime}&\equiv {L(\gamma_{1}\gamma^\prime_{2})=2\cosh^{-1}\left[\left|\frac{(-c^\prime_{1}\mu^{-1}+c^\prime_{2}\mu)}{2\sqrt{R^\prime_{1}R^\prime_{2}}}\right|\right]}. 
\end{align}

\section{Holographic Decoherence}
\label{sec:decohe}
Now we are ready to model an explicit model of holographic decoherence by using the four-boundary solutions constructed above. Remind that we were interested in a holographic setup discussed in the very first of section \ref{sec:wh} (see also Figure \ref{fig:setup}). We first discuss an explicit example in section \ref{subsec:amodel} and then move to a more general argument \ref{subsec:general}.  The discussion in this section is very similar to the one found in \cite{Balasubramanian:2020hfs,Li:2020ceg} where the entanglement between a black hole and Hawking radiation was studied using such multiboundary wormholes.

\subsection{A Model}\label{subsec:amodel}
We use the four-boundary wormhole constructed in the previous section. 
For simplicity, we always assume an $\mathbb{Z}_2$ symmetry, $c_i=-c_i^\prime$ and $R_i=R_i^\prime$ so that we can focus on the half of the whole system, say $AA^\prime$, and $AB$.

First, we start a thermo-field double state on $AB$, which is dual to an eternal BTZ black hole. We then attach these two systems $A$ and $B$ to two heat bathes/environments, $A^\prime$ and $B^\prime$. In this paper, we implicitly assume that the dimensions of the Hilbert space of the heat bathes/environments, $A^\prime$ and $B^\prime$ are the same order as the one of $A$ and $B$. The total system $ABA^\prime B^\prime$ has a holographic description by a four boundary wormhole. See Figure \ref{fig:setup}.  

Once the heat bathes/environments $A^\prime$ (and $B^\prime$) are attached to the original system $A$ (and $B$), an energy current goes between $A$ (and $B$), and $A^\prime$  (and $B^\prime$), which induces black hole mass change. 
Since the mass $M$ of BTZ black hole is related to the horizon length $L$ as $L=2\pi\sqrt{8G_NM}$, 
we model this energy current as following time evolution, 
\begin{align}
 S_{A}=\frac{L_{A}}{4G_{N}}=\frac{\sqrt{M_{0A}-\alpha t} }{4G_{N}}, \quad S_{A^\prime} =\frac{L_{A^\prime}}{4G_{N}}=\frac{\sqrt{M_{0A^\prime}+\alpha t} }{4G_{N}},  
\label{timeevoS}
\end{align}
where $\alpha$ is a parameter we set as a velocity of the energy flow\footnote{Here we abused the notation. We must not regard $t$ in \eqref{timeevoS} as the time coordinate appeared in the metric \eqref{eq:metric}. As discussed below, here we treat ``time'' $t$ as an order of snapshot solutions on a fixed time slice. }. $M_{0A}, M_{0A^\prime}$ imply initial values of mass parameters. Here we implicitly absorbed $8(2\pi)^2G_N$ into the definition of mass and velocity parameters for simplicity in such a way that BTZ horizon length $L$ and its mass $M$ is related as $L = \sqrt{M}$. $\mathbb{Z}_2$ symmetry implies the same for $B$ and $B'$.  
In this way, the conservation law of total energy in $A$ and $A'$ (and also $B$ and $B'$) is satisfied.  
Note that, since an energy current goes to reach equilibrium, we restrict our time evolution in the following range of time; 
\begin{align}
M_{0A}-\alpha t \ge M_{0A^\prime}+\alpha t  \quad  \Leftrightarrow \quad  0 \le t \le \frac{  M_{0A} - M_{0A^\prime}}{2 \alpha}   \,.
\label{timerange}
\end{align}
We model this time-evolution as a bunch of snapshot of the wormhole, where throats become smaller/larger as moduli changes.  
This is our time evolution model of the decoherence process, where through the energy flow between $A$ and bath $A'$ (and similarly $B$ and $B'$), system $A$ get entangled and becomes equilibrium with bath/environment $A'$.

Since our primary interest is how 
the amount of correlation, between $A$ and bath $A'$ and also between $A$ and $B$ evolves through the time evolution of decoherence,  
we need to evaluate the mutual information of between them.    
For that purpose, we need to identify how the moduli parameters of four boundary wormhole changes by the time evolution of our decoherence process.  
Under the $\mathbb{Z}_2$ symmetry, we have 5 moduli parameters; $\mu$, $c_1$, $c_2$, $R_1$, $R_2$. 
First, the length of horizon obeys the formula discussed in the previous section:
\begin{align}
\label{const1}
\cosh \dfrac{L_{A}}{2} &= \frac{ c_{1}-c_{2} }{2\sqrt{R_{1}R_{2}}} = {\cosh \left( \frac{\sqrt{M_{0A}-\alpha t}  }{2} \right)}, \\ 
\cosh \dfrac{L_{A^\prime}}{2} &= \frac{ c_{1}\mu^{-1}-c_{2}\mu}{2\sqrt{R_{1}R_{2}}} 
= {\cosh \left( \frac{\sqrt{M_{0A^\prime}+\alpha t} }{2} \right)} .
\label{const2}
\end{align}
In above second equalities, we use our time evolution model eq.~\eqref{timeevoS}. 
Second, as is seen in Figure \ref{fig:4bdry}, the 5 moduli should obey following geometric inequalities 
\be
1 < c_{2}-R_{2}, \quad c_{2}+R_{2} <c_{1}-R_{1}, \quad c_{1}+R_{1} <  \mu^2,  \label{eq: consistent}
\ee 
for consistency. This can be satisfied by introducing the following positive function $g > 0$; 
\begin{align}
1 + g = c_{2} - R_{2}, \quad c_{2}+R_{2} + g  = c_{1}-R_{1}, \quad c_{1}+R_{1} +g =  \mu^2, 
\label{eq:constg}
\end{align}
In other words, we tune moduli of wormholes by hand as \eqref{eq:constg}\footnote{Generically, \eqref{eq: consistent} can be satisfied by 
using 3 unknown positive functions $g_i > 0 $ ($i=1,2,3$); 
$1 + g_1= c_{2} - R_{2}$, 
$c_{2}+R_{2} + g_2  = c_{1}-R_{1}$,  
$c_{1}+R_{1} +g_3 = \mu^2 $.  
However since we do not know its full CFT description, just for simplicity in this paper, we set all $g_i$ the same function $g$.}.  

What is the appropriate function for $g(t)$? 
To choose an appropriate one, let us discuss what we expect by the endpoint of the decoherence process.   
From our assumption that the dimension of the Hilbert space of the heat bathes/environments $A^\prime$ is the same order as the one of $A$, at the final stage of the energy flow between $A$ and $A^\prime$, they are expected to reach equilibrium.  Therefore their masses become the same, and so are their areas;  
\begin{align}
L_{A} \to L_{A^\prime}  \quad {\mbox{at}} \quad  t \to  \frac{  M_{0A} - M_{0A^\prime}}{2 \alpha}  \,.
\end{align}
Then, from eq.~\eqref{const1} and \eqref{const2}, this implies that we need 
\begin{align}
\mu \to 1 \quad {\mbox{at}} \quad  t \to  \frac{  M_{0A} - M_{0A^\prime}}{2 \alpha}  \,,
\end{align} 
at the end point of the decoherence. 
This forces following constraints 
$c_{1}, c_{2} \rightarrow 1 \,,   R_{1}, R_{2} \rightarrow 0$ from the geometric constraint of Figure \ref{fig:4bdry} or eq.~\eqref{eq: consistent}, and this is possible {\it if and only if} $g \to 0$ at the end point of the decoherence. 
Therefore we impose 
\begin{align}
&g(t) >0 \,, \; \quad \frac{d g}{dt} <0 \,, \\
\mbox{and} \; \quad & g \to 0   \quad {\mbox{at}} \quad  t \to  \frac{  M_{0A} - M_{0A^\prime}}{2 \alpha}  \,
\label{eq:evog}
\end{align} 
as an appropriate function for $g(t)$. More concretely, we will choose following function for $g(t)$,  
\begin{align}
\label{eq:choiceofgt}
g(t) &=\epsilon(t) \times \left( 1+\mathrm{e}^{-\alpha t} \right) \, 
\end{align}
where  $\epsilon(t)$ is smooth monotonically decreasing function satisfying followings, 
\begin{align}
\label{epcondition1}
\epsilon(t) & \sim 1 \quad \mbox{when}  \quad t \ll  \frac{  M_{0A} - M_{0A^\prime}}{2 \alpha}  \,,  \\
\mbox{and} \quad \epsilon(t) & \to 0  \quad \,\, \mbox{as}   \,\,\,\, \quad t \to  \frac{  M_{0A} - M_{0A^\prime}}{2 \alpha} \,.
\label{epcondition2}
\end{align}
we seek the time evolution of the moduli parameters under this.

Now our task is to determine the time evolution of the 5 moduli parameters $\mu$, $c_1$, $c_2$, $R_1$, $R_2$, by solving 5 constraint equations \eqref{const1}, \eqref{const2}, \eqref{eq:constg}, with an appropriate choice of $g$ satisfying \eqref{eq:evog}. With that, we evaluate the time evolution of the mutual information between $A$ and $A^\prime$ and also between $A$ and $B$. 
For that purpose, let us solve the constraint equations \eqref{eq:constg} for $c_1$, $R_1$ and $R_2$ as,   
\be
\label{eq:c1r1r2}
c_{1} =c_{2} + \frac{\mu^2-1-g}{2} , \quad R_{1} =-c_{2} +\frac{1+ \mu^2-g}{2},  \quad R_{2} =c_{2} - 1 - g .  
\ee
Then, by plugging these equation into \eqref{const1} and \eqref{const2}, we obtain equations for $\mu$ and $c_{2}$ as follows
\be 
\cosh \frac{L_A}{2} =\frac{   \mu^2 -1-g }{4   \sqrt{\left(- c_{2} +\frac{1+   \mu^2 -g}{2} \right)  (c_{2} -1-g)}}, 
\label{eq:L1} 
\ee
and 
\be 
\mu\frac{\cosh \frac{L_{A^\prime}}{2}}{\cosh \frac{L_{A}}{2}} = {\left|\dfrac{-2c_2(\mu^2-1)+   \mu^2  -1 -g}{-   \mu^2 +1 + g}\right|}  \,. 
\label{eq:equilibium}
\ee
We proceed as follows;  
for given $g(t)$, we first solve the equations \eqref{eq:L1} and \eqref{eq:equilibium} numerically to determine $c_2$ and $\mu$, then  $c_1$, $R_1$, $R_2$ are determined from \eqref{eq:c1r1r2}. In this way, all time-dependence of the moduli are specified. Then from these moduli, 
we can plot the mutual information between $A$ and $A^\prime$, $I(A:A^\prime)$ as a function of time, where $I(A:A^\prime)$ is 
\begin{align}
I(A:A^\prime) &=S_{A} +S_{A^\prime} -S_{AA^\prime} \,,  
\label{eq:mutuals} \\
S_{AA^\prime} &=\min \left[S_{A}+S_{A^\prime}, \, \frac{L_1}{4G_{N}} \right] \,, \quad 
\label{eq:SAAp}
L_1=2\log\mu.
\end{align}
We will call it a disconnected phase when the relationship $S_{AA^\prime}=S_{A}+S_{A^\prime}$ is satisfied, and a connected phase when the relationship $S_{AA^\prime}=L_1/4G_{N}$ is satisfied.

\begin{figure}[t]
 \begin{center}
 \hspace{-5mm}
\resizebox{130mm}{!}{\includegraphics{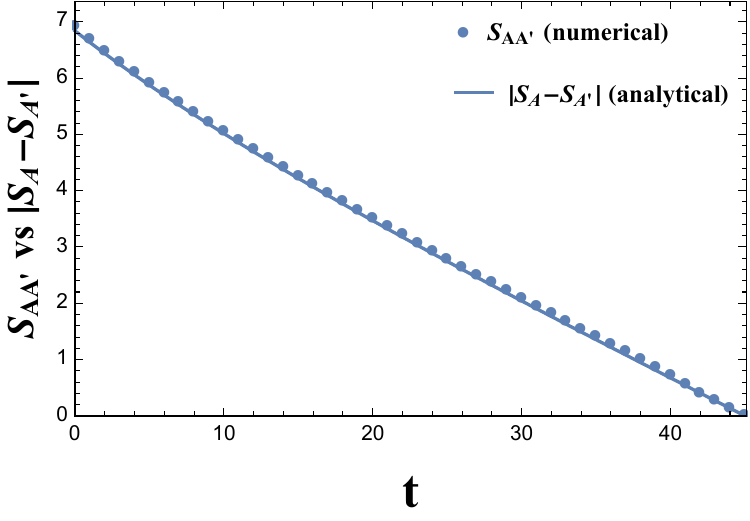}} 
 \end{center}
 \vspace{-4mm}
 \caption{A numerical plot of $S_{AA^\prime}$ (blue dots) and an analytical plot of $|S_{A}(t)-S_{A^\prime}(t)|$ (blue line) 
 for the choice of $g(t)$ given by eq.~\eqref{eq:choiceofgt}, \eqref{eq:functionepsilon}. 
  where $\epsilon_0 =1.0\times10^{-2}$. 
  As an initial condition, we took $M_{0A} =100 , M_{0A^\prime} = 10 $, with $\alpha = 1$. 
  In this plot, we set $4G_N=1$.
 }
 \label{fig:plots}
\end{figure}

\begin{figure}[t]
 \begin{center}
\resizebox{115mm}{!}{\includegraphics{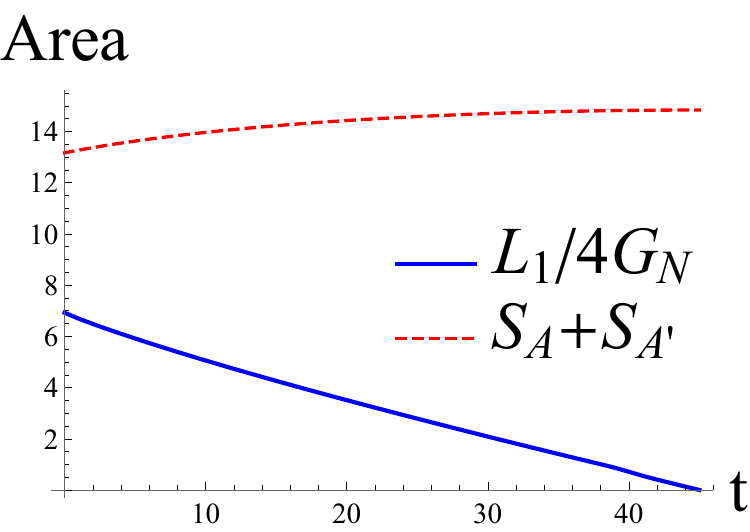}}
 \end{center}
 \caption{A numerical plot of $S_{AA^\prime} = L_1/4G_N$ (blue line) and an analytical plot of $S_{A}(t) + S_{A^\prime}(t)$ (red dashed line) with the same choice of parameters as Figure \ref{fig:plots}. Clearly at the end of the decoherence process at $t=45$, $L_1$ goes to zero.  For the illustration of the geometry, see Figure \ref{fig:LALApL1} as well. 
}
 \label{fig:plot_area}
\end{figure}

We have solved numerically the time-dependence of the moduli for the choice of 
$M_{0A}= 100$, 
$M_{0A^\prime}=10$,  and $\alpha=1$. 
Mass equilibrium is reached at 
\begin{align}
t = \frac{M_{0A} - M_{0A^\prime}}{2 \alpha}= 45.
\end{align} 
$g(t)$ is given by eq.~\eqref{eq:choiceofgt}, {\it i.e.,} $g(t)=\epsilon(t) \times \left( 1+\mathrm{e}^{-\alpha t} \right)$.
Given the mass parameters, our choice for smooth function $\epsilon(t)$ is 
\begin{align}
\label{eq:functionepsilon}
\epsilon= \epsilon_0 \frac{1 - \tanh \left( t - 40 \right) }{2} 
\end{align}
with $\epsilon_0 = 1.0 \times10^{-2}$. 
This satisfies eq.~\eqref{epcondition1} and \eqref{epcondition2} in good accuracy\footnote{In fact, one can confirm that our results are unaffected as long as we choose any $\epsilon(t)$, satisfying eq.~\eqref{epcondition1} and \eqref{epcondition2}.}. 
We also set $4 G_N = 1$. The results are shown in Figure \ref{fig:plots} and Figure \ref{fig:plot_area}.  

\begin{figure}[t]
 \begin{center}
  \resizebox{55mm}{!}{\includegraphics{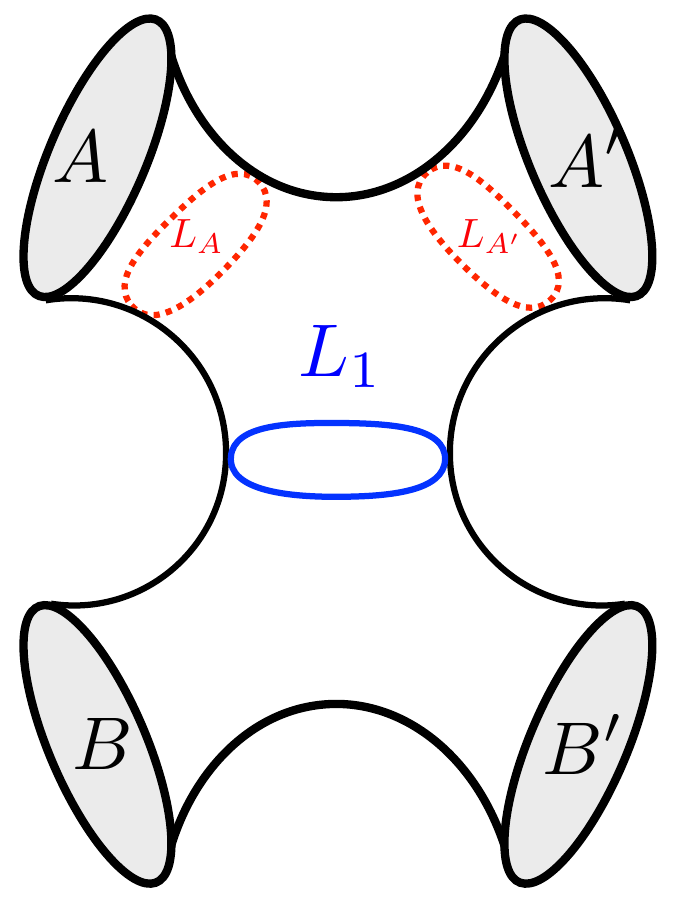}}
 \end{center}
 \caption{{Four boundary wormhole. The ``areas'' of the neck region $L_A$, $L_{A^\prime}$, and $L_1$ are shown pictorially. As decoherence process evolves,  $L_1$ shrinks to zero.}}
 \label{fig:LALApL1}
\end{figure}

Some comments are in order; 
\begin{enumerate}
\item The von-Neumann entropy always satisfy so-called Araki-Lieb inequality\footnote{We thank an anonymous referee for pointing out the related discussion which improves our previous numerical results.}, 
\be
{|S_A - S_{A^\prime}|\leq S_{AA^\prime}\leq S_A + S_{A^\prime}.}
\ee
In our context, the upper bound corresponds to the disconnected phase. Note that during the decoherence process, $S_A > 0$ and $S_{A^\prime} > 0$ hold. Then our numerical solution shows that the lower bound of the Araki-Lieb inequality,  
{\it i.e.}, 
\begin{align}
 |S_A - S_{A^\prime}| = S_{AA^\prime} <  S_A + S_{A^\prime}.
\end{align}
is always saturated during the decoherence process. 
\item Therefore our numerical solution always show the connected phase during the decoherence process and 
\be
S_{AA^\prime}=\frac{L_1}{4G_{N}}  = \frac{\log \mu}{2 G_N}
\ee
always holds during the process. At the end of the decoherence process at $t \to 45$, $S_{AA^\prime}(t) \to 0$,  therefore $L_1 \to 0$ ($\mu \to 1$) and the bulk of a four-boundary wormhole pinches off. See Figure \ref{fig:LALApL1}. 
\item This parameter range of time $t$ is consistent with \eqref{timerange}, where we have $({  M_{0A} - M_{0A^\prime}})/{2 \alpha} = 45 $. 
\item We have also found similar behavior for different parameter regimes.  
\end{enumerate}
Note that even though in the limit $\mu \to 1$, we have $c_{1}, c_{2} \rightarrow 1 \,,   R_{1}, R_{2} \rightarrow 0$  and therefore the inner and outer blue semi-circle in Figure \ref{fig:4bdry} collapses and coincides, in such a limit the area of $A$ for $S_A$ and $A^\prime$ for $S_{A^\prime}$ does not vanish as is seen for $S_A + S_{A^\prime}$ in Figure \ref{fig:plot_area}.   
This is because  
Figure \ref{fig:4bdry} omits the warped factor ${|\Im Z|^{-2}}$ of the metric \eqref{warping}. Therefore the enhancement by this warped factor makes the area finite. The resulting four boundary wormhole is drawn in Figure \ref{fig:LALApL1}.

Once we confirm the time-evolution to $\mu \to 1 \Leftrightarrow L_1 \to 0$ and therefore to $S_{AA^\prime} \to 0$ exists, then we can see that there is no correlation between A and B, both classically and also quantum mechanically. This can be understood as follows; 
since the whole system $AA^\prime B B^\prime$ is pure state and due to $\mathbb{Z}_2$ symmetry,  
$S_{AA^\prime} \to 0$ implies that the mutual information between $AA^\prime$ and $B B^\prime$ is zero, 
\be
I(AA^\prime : BB^\prime) = S_{A A^\prime} +S_{B B^\prime} -S_{A A^\prime B B^\prime} = 0 \,, 
\ee
namely, the final state should possess a product structure $\rho_{AA^\prime} \otimes \rho_{BB^\prime}$, where $\rho_{AA^\prime}$ ($\rho_{BB^\prime}$)  represents the density matrix of ${AA^\prime}$ (${BB^\prime}$). 
Due to strong subadditivity inequality, the mutual information never increases by 
tracing out subsystems, $A^\prime$ and $B^\prime$ in this case. Therefore we conclude that 
\be
I(AA^\prime : BB^\prime) \ge I(A : B) =  0  \,.  
\ee
This means that the original entanglement between $A$ and $B$ completely disappears, both classically and also quantum mechanically.

\subsection{Absence of correlation at the end point}\label{subsec:general}
One can show that, under the mild assumptions, the end point of  {\it any holographic} decoherence process cannot have correlations either classically and also quantum mechanically.  
The argument is closely related to the absence of  holographic duals of GHZ type states \cite{Hayden:2011ag}.  

For that purpose, we need following two assumptions at the end point of the holographic decoherence processes: 
\begin{enumerate}
\item  The entropies of each subsystem are  identical, 
\be
\label{assump1}
S_{A} =S_{A^\prime} =S_{B}=S_{B^\prime} \,.
\ee 
This is motivated by the thermal equilibrium of the end point.  
\item  The mutual information between the system and the bath/environment, {\it i.e.},  $A$ and $A^\prime$ saturates the bound, 
\be
\label{assump2}
I(A:A^\prime)\leq S_{A} +S_{A^\prime}, 
\ee 
namely, $S_{AA^\prime}=0$. This also implies $I(B:B^\prime)$ saturates the bound, since  $S_{AA^\prime}=S_{BB^\prime}$. 
\end{enumerate}

Because the mutual information is monotonic, we have\footnote{We thank Zixia Wei for discussion on this. } 
\begin{align}
I (AA^\prime: B)&\ge I(A:B) ,\\
I (AA^\prime: B)&\ge I(A^\prime:B).
\end{align}
Each left hand side $ I (AA^\prime: B) = S_{AA^\prime} + S_{B} - S_{AA^\prime B} $ is vanishing, due to the conditions $S_{AA^\prime} =0, S_{AA^\prime B} =S_{B^\prime}$. This implies $  I(A:B) = I(A^\prime:B)=0$, {\it i.e.}, 
{there is no classical and quantum correlations between two systems $A$ and $B$.} This is exactly what we have seen in our decoherence model of the previous subsection. 

Since our assumptions \eqref{assump1} and \eqref{assump2} are quite natural, we conjecture that under the decoherence process which admits holographic duals,  
no classical and quantum correlations between two systems. We will discuss this issues more in the discussion section. 

\subsection{Loopholes}
There are several loopholes in this argument. 
\begin{enumerate}
\item We assume that the dimensions of the Hilbert space of $A$ and $A^\prime$, and through the $\mathbb{Z}_2$ symmetry the ones of $B$ and $B^\prime$, are the same, therefore at the equilibrium, their entropy is the same as \eqref{assump1}. Obviously if we relax this condition, then the argument of this subsection breaks down and there is no guarantee that under the decoherence process, one obtain $I(A^\prime:B)=0$. 
\item Another loophole of the discussion in this section is that even though there is no classical and quantum correlations between two systems, if we consider only the the low energy subspace (so-called ``code subspace''), then there is a possibility that in that subspace, one might see the state as if it admits some correlations. 
\end{enumerate} 
Regarding the second possibility, in fact, in \cite{Verlinde:2020upt} Verlinde proposed an interesting connection between a particular separable state, dubbed thermo mixed double (TMD) state, and the thermo-field double (TFD) state in the light of the low energy subspace (so-called code subspace). 
They are defined in bi-partite system as follows;
\begin{align}
\ket{\textrm{TFD}}&=\sum_{n} \sqrt{p_n}\ket{nn},\\
\rho_{\textrm{TMD}}&=\sum_{n} p_n\ket{nn}\hspace{-1mm}\bra{nn},
\end{align}
where $p_n$ is the standard Boltzmann factor $p_n=\mathrm{e}^{-\beta E_n}/Z(\beta)$.
Obviously $\ket{\textrm{TFD}}$ is a pure state and $\rho_{\textrm{TMD}}$ is a mixed state. The former has quantum correlation (entanglement) and the latter has only classical correlation. 
Note that these both states are particular extensions of thermal state. 
Verlinde \cite{Verlinde:2020upt}  pointed out that for a low energy observer (code subspace), it might be impossible to distinguish the two, namely, 
\be
P_{\textrm{code}}\,\rho_{\textrm{TMD}}\,P_{\textrm{code}}=P_{\textrm{code}}\ket{\textrm{TFD}}\hspace{-1mm}\bra{\textrm{TFD}}P_{\textrm{code}},
\label{eq:verlinde}
\ee
where $P_{\textrm{code}}$ is a projection operator onto the code subspace. We discuss this possibility in next section  in detail.

\section{Thermo mixed double state and code subspace}
\label{sec:toy}

As we have seen in previous section, in the full space, one can distinguish these two states, thermo mixed double (TMD) state and the thermo-field double (TFD) state\footnote{It is also clear from various measures for mixed states conjectured to be related to the entanglement wedge cross-section\cite{Takayanagi:2017knl,Nguyen:2017yqw,Kudler-Flam:2018qjo,Tamaoka:2018ned,Dutta:2019gen}. For example, the logarithmic negativity\cite{Vidal:2002zz} and the (regularized) odd entropy\cite{Tamaoka:2018ned} are clearly zero for the TMD state as $\rho_{\textrm{TMD}}$ is invariant under the partial transposition. This kind of indistinguishablity in CFT was studied using relative entropy in \cite{Takayanagi:2018zqx}.}, and we have seen explicitly that no correlation are left after the decoherence both classically and also quantum mechanically in our holographic model. However if we restrict only to a code subspace (low-energy) observer, this might be possible. We now discuss this possibility.  
An important remark is that the argument  in \cite{Verlinde:2020upt} is based on the perturbation theory in Newton constant $G_N$. 
In other words, the above relation is true only {up to non-perturbative corrections}. 

In this section, we would like to clarify a possible (in)distinguishability of TFD and TMD states by utilizing the local operator correlation functions. 
This illustrates what is necessary to distinguish TFD and TMD states. 
Here, we implicitly assume the system is chaotic enough and the eigenstate thermalization hypothesis (ETH) \cite{Srednicki} is satisfied for local operators. The ETH demands local operators $\mathcal{A}=\mathcal{O}_L, \tilde{\mathcal{O}}_R$ obey the following matrix elements,
\begin{equation}
\braket{E_m|\mathcal{A}|E_n}=\delta_{mn}\bar{A}(\bar{E}_{mn})+e^{-S(\bar{E}_{mn})/2}R_{mn}f_{\mathcal{A}}(\bar{E}_{mn},\omega_{mn}),\label{eq:eth}
\end{equation}
where $\bar{E}_{mn}=(E_m+E_n)/2$, $\omega_{mn}=(E_m-E_n)/2$. Here $\bar{A}(\bar{E}_{mn}), f_{\mathcal{A}}(\bar{E}_{mn},\omega_{mn})$ are the smooth functions depend on the given operator $\mathcal{A}$. In the second term, $S(\bar{E}_{mn})$ is the thermodynamic entropy at energy $\bar{E}_{mn}$ and $R_{mn}$ is a random distribution with zero mean and unit variance. 

\subsection{TMD vs TFD - round 1: two-point function}\label{sec:eth}
To discuss the difference between TMD and TFD, we study the two-point correlation function of certain operators,
\begin{align}
\braket{\mathcal{O}_L(t_L)\tilde{\mathcal{O}}_R(t_R)}_{\textrm{TMD}}&\equiv\mathrm{Tr}_{LR}(\rho_{\textrm{TMD}}\mathcal{O}_L(t_L)\tilde{\mathcal{O}}_R(t_R)), \label{eq:expTMD}\\
\braket{\mathcal{O}_L(t_L)\tilde{\mathcal{O}}_R(t_R)}_{\textrm{TFD}}&\equiv\mathrm{Tr}_{LR}(\ket{\textrm{TFD}}\bra{\textrm{TFD}}\mathcal{O}_L(t_L)\tilde{\mathcal{O}}_R(t_R)) \label{eq:expTFD}.
\end{align}
Discussion in this subsection needs ETH, but other than that, it is generic and does not need to assume holographic CFT. 
First of all, even if we focus on such low-energy operators, we can see that \eqref{eq:expTMD} and \eqref{eq:expTFD} behave differently. 
In particular, the expectation value for the TMD state \eqref{eq:expTMD} has no time-dependence,
\begin{align}
\braket{\mathcal{O}_L(t_L)\tilde{\mathcal{O}}_R(t_R)}_{\textrm{TMD}}&=\sum_{n}p_n\braket{n|\mathcal{O}_L(t_L)|n}\braket{n| \tilde{\mathcal{O}}_R(t_R)|n}\nn\\
&=\sum_{n}p_n\braket{n|\mathcal{O}_L(0)|n}\braket{n|\tilde{\mathcal{O}}_R(0)|n}\nn\\
&=\braket{\mathcal{O}_L(0)\tilde{\mathcal{O}}_R(0)}_{\textrm{TMD}}. \label{eq:notime}
\end{align}
On the other hand, the one for the TFD state \eqref{eq:expTFD} can be used to probe the black hole interior \cite{Maldacena:2001kr,Israel:1976ur,Balasubramanian:1998de}.  
This difference obviously comes from the contribution from {\it off-diagonal elements} in the TFD state. Notice that these elements are directly related to entanglement of the TFD state between $L$ and $R$.  

Given that off-diagonal elements manifestly distinguish the two, 
one possible way to identify $\eqref{eq:expTMD}$ with $\eqref{eq:expTFD}$ is to take the time average and smear out the off-diagonal elements. 
For example, the standard time average gives the exact agreement, 
\be
\lim_{T\rightarrow\infty}\dfrac{1}{T}\int^{T}_{-T}dt\braket{\mathcal{O}_L(t)\tilde{\mathcal{O}}_R(0)}_{\textrm{TMD}}=\lim_{T\rightarrow\infty}\dfrac{1}{T}\int^{T}_{-T}dt\braket{\mathcal{O}_L(t)\tilde{\mathcal{O}}_R(0)}_{\textrm{TFD}}.
\ee
because the time-average of the off-diagonal elements in ETH gives zero. 
In \cite{Verlinde:2020upt}, it is argued that a more discrete average based on the Poincare recurrence time should be possible.  
Note that, however, in any cases, the \eqref{eq:notime} suggests that we cannot see the growth of Einstein-Rosen bridge by using such correlation functions, which is manifest difference between TMD vs TFD.

\subsection{TMD vs TFD - round 2: four-point function}
Next, we would like to evaluate the following four point function 
\be
I_{4}(x)={\rm Tr } \left[ \rho_{\textrm{TMD}} \mathcal{O}_{L}(x) \mathcal{O}_{L}(1) \tilde{\mathcal{O}}_{R}(x) \tilde{\mathcal{O}}_{R}(1)\right],
\ee
for the TMD state in the holographic CFT.  
One can write this as follows; 
\be 
I_{4}(x)=\int dh \; d \bar{h} \; \rho (h,\bar{h}) \; P(h,\bar{h}) \left[\langle h, \bar{h}|\mathcal{O}_{L}(x) \mathcal{O}_{L}(1)|h, \bar{h} \rangle  \right]  \left[\langle h, \bar{h}| \tilde{\mathcal{O}}_{R}(x) \tilde{\mathcal{O}}_{R}(1)|h, \bar{h} \rangle  \right] . 
\ee
As is clear in this example, this four-point function essentially has the same structure as two-point function seen in previous subsection; if replace $\mathcal{O}_L(t_L)$  of \eqref{eq:notime} as  $\mathcal{O}_{L}(x) \mathcal{O}_{L}(1)$, and so is for $\tilde{\mathcal{O}}_R(t_R)$, we obtain this expression.   
The integral naturally factorized into chiral and anti-chiral parts $I_{4}=I \bar{I}$ due to the factorizaiton 
\begin{equation}
  \rho(h,\bar{h}) =\rho(h) \rho(\bar{h}) \quad P(h,\bar{h})=P(h) P(\bar{h}) \,, 
\end{equation}
where 
\begin{equation}
    \rho(h)= \exp \left[2\pi \sqrt{ \frac{c h}{6}}\right], \quad P(h)=\exp[-\beta h] \,.
\end{equation}

An explicit form of $I$ is given by
\be
I= \int dh \rho(h)P(h) \exp[-\frac{c}{3} g(h, \varepsilon,z)] \equiv \int dh e^{K(h,z)},
\ee
where we assume the vacuum contribution is dominant\cite{Hartman:2013mia}\footnote{For simplicity, we tuned the value of $x$ so that another possible channel\cite{Kusuki:2019rbk,Kusuki:2019evw} is not dominant.} and use the ‘heavy-heavy-light-light’ conformal block\cite{Asplund:2014coa},
\be
\langle h |\mathcal{O}(x)\mathcal{O}(1)| h \rangle =\exp[-\frac{c}{6} g(h, \varepsilon,z)-\frac{c}{6}g(h, \varepsilon,\bar{z})]
\ee
with 
\be 
g(h, \varepsilon;1-z)= \frac{\varepsilon}{2} \log \left(\frac{1-z^{\alpha}}{\alpha} \right)+\frac{\varepsilon}{4}\log z,
\ee
where
\be 
\alpha=\sqrt{1-\frac{24h}{c}}. 
\ee

We can evaluate this integral by a saddle point approximation once assuming the saddle $h_{*}$ is $O(c)$, $h=\gamma c$. In this case $K(h,z)=cL(\gamma,z)$ and 
\be 
L(\gamma,z)= 2\pi \sqrt{\frac{\gamma}{6}}-\beta \gamma -\frac{1}{6} g(\gamma, \varepsilon ,z) \,.
\ee
The saddle $\gamma=\gamma_{*}$ satisfies 
\be 
\frac{\partial L}{\partial \gamma} =0 \,.
\ee
Because $\varepsilon \ll 1$, $g(\gamma, \varepsilon ,z) \ll1$, therefore the saddle is approximated by 
\be 
\gamma_{*}=6 \left(\frac{2\pi}{\beta} \right)^{2} \,.
\ee
This just usual thermal temperature-energy relation, which is plausible. 
This means, in the large $c$ limit where one can use saddle point approximation, the four point function always gets factorized into the 2 point functions, 
\be 
{\rm Tr} \left[ \rho_{\textrm{TMD}} \mathcal{O}_{L}(x) \mathcal{O}_{L}(1) \mathcal{O}_{R}(x) \mathcal{O}_{R}(1)\right]= \left({\rm Tr}  \left[ e^{-\beta H} \mathcal{O}_{L}(x) \mathcal{O}_{L}(1) \right]\right)^{2}
\ee

On the other hand the four point function of the thermo-field double state does not get factorized due to the off-diagonal elements. This difference results in the difference of the bulk causal structures \cite{Maldacena:2001kr,Israel:1976ur,Balasubramanian:1998de}, in particular the existence of the horizon interior. 

Given these, if we define code subspace as low-energy observers who are blind to the off-diagonal elements, then in code subspace, it is not possible to distinguish the difference.

\subsection{TMD vs TFD - round 3: code subspace toy model}
Since off-diagonal elements give the difference between thermo mixed double (TMD) and thermo-field double (TFD) state,  if one defines code subspace as a low energy observer who is blind to the off-diagonal elements, then one can also construct a simple toy model, satisfying \eqref{eq:verlinde}, and furthermore still maintain low energy entanglement. In this subsection, we provide such a toy model example. 

Let us consider the two spin systems, say spin $A$ and $B$, whose Hamiltonian is given by
\be
H=  - N^2 \, \bold{J}^2_{AB} = - \frac{R^2}{G_N} \, \bold{J}^2_{AB},
\ee
where $N^2 =  \frac{R^2}{G_N}$ is some large number and $\bold{J}^2_{AB}$ is the square of the total angular momentum, namely the Casimir operator for $SU(2)$. 
We have two energy levels $E_0<E_1$ whose eigenvectors are generically given by
\begin{align}
\ket{0_{AB}}&=\mathcal{N}^{-\frac{1}{2}}\bigg[\alpha_{\uparrow}   \ket{\uparrow_A\uparrow_B}+\alpha_* \big(\ket{\uparrow_A\downarrow_B}+\ket{\downarrow_A\uparrow_B}\big)+\alpha_{\downarrow} \ket{\downarrow_A\downarrow_B}\bigg], \\
\ket{1_{AB}}&=\dfrac{1}{\sqrt{2}}\big(\ket{\uparrow_A\downarrow_B}-\ket{\downarrow_A\uparrow_B}\big),
\end{align}
where $\alpha_\uparrow, \alpha_*, \alpha_{\downarrow}$ are arbitrary constants and $\mathcal{N}$ is a normalization factor. 
Note that our ground states degenerate. 

In this Hamiltonian, ``Gibbs ensemble'' is given by
\be
\rho(\beta)=\dfrac{1}{Z(\beta)}(3e^{-\beta E_0}\ket{0_{AB}}\hspace{-1mm}\bra{0_{AB}}+e^{-\beta E_1}\ket{1_{AB}}\hspace{-1mm}\bra{1_{AB}}),
\ee
thus, the corresponding thermo-field double state should have the following form,
\be
\ket{\textrm{TFD}}_{ABA^\ast B^\ast}=\dfrac{1}{\sqrt{Z(\beta)}}\left(\sqrt{3}e^{-\beta E_0/2}\ket{0_{AB}0_{A^\ast B^\ast}}+e^{-\beta E_1/2}\ket{1_{AB}1_{A^\ast B^\ast}}\right).
\ee
On the other hand, the thermo mixed double state is given by
\be
\rho_{\textrm{TMD}}=\dfrac{3e^{-\beta E_0}}{Z(\beta)}\ket{0_{AB}0_{A^\ast B^\ast}}\hspace{-1mm}\bra{0_{AB}0_{A^\ast B^\ast}}+\dfrac{e^{-\beta E_1}}{Z(\beta)}\ket{1_{AB}1_{A^\ast B^\ast}}\hspace{-1mm}\bra{1_{AB}1_{A^\ast B^\ast}}.
\ee

Suppose one take the $G_N \to 0$ limit. Then a natural low energy code subspace is the space with energy $E_0$. This is because a excited state with energy $E_1$ is suppressed by the exponential factor 
\be
\sim e^{-  \frac{\beta R^2}{G_N} (E_1 - E_0)} \ll 1
\ee
which is tiny in the large $N$, equivalently $G_N \to 0 $ limit. Therefore after projection onto the low energy (=ground state) code subspace, one can see that \eqref{eq:verlinde} is satisfied in this model.  
Note that due to the degeneracies of the ground states, {\it i.e.,} $\dim\mathcal{H}_{\textrm{code}}>1$, even in the code subspace, 
we can maintain certain amount of entanglement. 
One can easily extend the above argument to the generalized thermo mixed double states discussed in \eqref{eq:verlinde}. 
It might be worth noting that the resulting entangled state has only $\mathcal{O}(N^0)$ degeneracy in the light of the AdS/CFT as we are looking at the low-energy code subspace.

\section{Conclusions and discussion}
\label{sec:disc}

The question  we addressed in this paper  is, to what extent  correlation  contained   in the boundary state  can affect the structure of the bulk spacetime in AdS/CFT correspondence. More concretely, we tried to find an Einstein Rosen bridge, which  is supported only by classical correlation of the boundary state \cite{Verlinde:2020upt}.  To construct this, we  considered  a decoherence process in which  an initial  thermo-field double state   evolves to an mixed state, due to interaction with environment.   In this process,  the quantum entanglement that the initial TFD state has  is eventually turned to classical correlation. In the holographic dual gravity description of  this process  involves an eternal BTZ black hole dual to the TFD state, attached to another  eternal black hole 
modeling the heat bath. The total system is dual to a four boundary wormhole,  with a particular entanglement structure.  We studied time evolution of  various bipartite  correlations in the system. We then find that  the final state 
can not have {\it any} correlation.  This implies  that we can not construct an ER bridge  dual to classical correlation in this way. We believe this is a distinguishing property of holographic theory, which is genuinely chaotic.  For example, in a generic quantum field  theory,  a decherence process whose final state contains classical correlation can be explicitly constructed as in \cite{delCampo:2019qdx}.
Although this non existence of correlation is a simple  observation in this particular  multiboundary wormhole setup,  we argued that  MHH inequality\cite{Hayden:2011ag}   prohibits  to  produce  such ER bridge  from decoherence in general in  holographic setup.

\vspace{0.2 cm}
We also discuss the distinguishability between thermo-field double and thermo mixed double state, which contains only classical correlations. Assuming  the eigenstate thermalization hypothesis (ETH), our conclusion is that as far as low energy observer are blind to the off-diagonal elements, then it is impossible to distinguish these two states. Furthermore we also construct a simple toy model which realizes this. An important aspect of our toy model is that it contains low energy degeneracy and therefore one can have entanglement structure even for low energy observer. 

\vspace{0.2 cm}

This conclusion can be test in simple quantum models  admitting  a gravity dual.  For example one can imagine  studying  this process in four copies of the  SYK model \cite{Sachdev:1992fk,Kit}.   First of all, out of two coupled SYKs, one can  construct a TFD state.  By preparing two  such TFD states in this way,  the decoherence process  can be concretely realized. We expect, due to highly chaotic nature of this model, the final state of the decoherence process in this model is  very similar to the one we obtained from the holographic point of view, namely that it does not contain any correlation even classically. It would be interesting to check this conjecture. Related discussion can be found, for example  in 
\cite{Almheiri:2019jqq,Piroli:2020dlx}.

\vspace{0.2 cm}

It would be also interesting to further study the time dependent gravity solution  dual to the decoherence.
Our  treatment of time evolution is heavily using the topological nature of three dimensional  pure gravity.  The initial time slice of  four boundary wormhole solution we dealt with,  depends on several moduli parameters  since it is a Riemann surface with four boundaries.  We  studied the holographic decoherence, by making  these moduli time dependent.  We saw this is still a solution of the Einstein equations  in three dimensions, and exhibits interesting dynamics.  
 In principle, we can construct a full fledged  time dependent spacetime, by preparing two  BTZ black holes, attaching  them, and following the time dependence by solving the equations of motion. In our analysis, we assumed a $\mathbb{Z}_2$ symmetry between subsystems $AA^\prime$ and $BB^\prime$ as we did not introduce interactions between them. Introducing interactions which break this $\mathbb{Z}_2$ symmetry would be interesting since traversable (multi-boundary) wormholes can realize. 

\vspace{0.2 cm}

Another avenue which deserves further investigation is, to understand better why it is hard to geometrize classical correlations, compared to quantum correlations. The fact that separable states in a bipartite system  
can not have gravity dual with smooth geometry is closely related, to the fact that a GHZ type state in a tripartite system  in CFT  side can not  be realized  by classical three boundary wormhole. This is because, such   separable states can be constructed by tracing out  one of the  Hilbert space of the tripartite system.

\section*{Acknowledgement} 
We would like to thank Yoshifumi Nakata and Zixia Wei for useful discussion.  We would also like to thank an anonymous referee for pointing out important errors in the previous version, which help us to improve our numerical analysis significantly. 
The work of N.I. was supported in part by JSPS KAKENHI Grant Number 18K03619.  
The work of K.T. is supported by the Simons Foundation through the ``It from Qubit'' collaboration. The work of K.T. is also supported in part by JSPS KAKENHI Grant Number 19K23441. T.U is supported by JSPS Grant-in-Aid for Young Scientists  19K14716. 
 

\appendix


\section{Review of known properties of \AdSt}
\label{AppA}
In this appendix, we review the properties of the $AdS_3$ geometry and its isometry, which we will use  to obtain wormholes as quotient.  See \cite{Skenderis:2009ju} for example. We use AdS length scale $\ell_{\textrm{AdS}}
=1 $ convention. 

\subsection{\AdSt} 
$AdS_3$ is defined by embedding it as 
\begin{align}
ds^2 &= -d U^2 - dV^2 + dX^2 + dY^2 \,, \\
& - U^2 - V^2 + X^2 + Y^2 = - 1  \,, \quad
\end{align}
in $\mathbb{R}^{2,2}$.

Poincar\'e coordinates are defined as 
\begin{align}
&U = \frac{1}{2z} \left( x^2 - t^2 + z^2 + 1 \right) \,, \quad V = \f{t}{z}  \,,  \nonumber\\
&X  = \frac{1}{2z} \left( x^2 - t^2 + z^2 - 1 \right) \,, \quad Y  = \f{x}{z} \,,
\label{Poincarecoord}
\end{align}
and we obtain the Poincar\'e metric 
\be
ds^2 = \frac{- dt^2 + dx^2 + dz^2}{z^2} \,.
\ee
Especially, $t=0$ slice in Poincar\'e coordinate is expressed as
\begin{align}
\label{tslicePoincare}
V=0 \,, \quad Y =\f{x}{z} \,, \quad 
U+X  = \f{x^2 + z^2}{z} \,, \quad 
U-X = \frac{1}{z} \,. 
\end{align}
Then, the $t=0$ slice in Poincar\'e coordinate (which is equivalent to $V=0$ slice) becomes hyperbolic space $H^2$ with its metric  
\be
ds_{t=0}^2 = \frac{dx^2 + dz^2}{z^2}  = \frac{dZ d\bar{Z}}{|\Im Z|^2} \,,
\label{warping}
\ee
In the second inequality, we define $Z \equiv x+ i z$.

 \subsection{Geodesic length in \AdSt}\label{App2}

Here we derive the length of geodesics in $AdS$ \cite{Bengtsson1998}.  
Let us define the vector $W^A$ in the embedding space $\mathbb{R}^{2,2}$, where
\begin{align}
W^A &= (U, V, X, Y) \,, \quad \eta_{AB} = \mbox{diagonal} 
\{ -1, -1, 1, 1 \} \\
W^2  &\equiv  W^A W^B \eta_{AB} =  - U^2 - V^2 + X^2 + Y^2 \,.
\end{align}
Geodesic equation can be derived from the following Lagrangian for the worldline, 
\be
L=\dfrac{1}{2} \dot{W}^A \dot{W}^B \eta_{AB} +\lambda (W^2+1) \,,
\ee
where $\tau$ is the affine parameter with $\dot{W}^A\equiv\dfrac{dW^A}{d\tau}$, and $\lambda $ is the Lagrange multiplier, to force trajectory on $AdS_3$. Then, the equation of motion is, 
\be
\ddot{W}^A=2 \lambda W^A \,, \label{EOMW}
\ee
and this yields 
\be
\label{geosolu}
 W^A(\tau) = c^A_+ e^{\sqrt{2 \lambda} \tau}   +  c^A_- e^{- \sqrt{2 \lambda} \tau} \,,
\ee
where $c^A_+, c^A_-$ are null and orthogonal constant vector in $\Real^{2,2}$, satisfying 
\be
\label{cconstraint}
c_+^A c_+^B \eta_{AB} = c_-^A c_-^B \eta_{AB}   =0,\;\;\;  2 c_+^A c_-^B \eta_{AB} =- 1 \,.
\ee

Note that from the Lagrange multiplier, we have 
\be
\label{Wconstraint}
W^2 = -1  \quad \Rightarrow  \quad W^A \dot{W}^B \eta_{AB} = 0 \,.
\ee
Taking the $\tau$ derivative twice of this equation, we see 
\be
\sqrt{ 2 \lambda}  =  \sqrt{\dot{W}^2}  = \mbox{constant} \,.
\ee
and that the geodesic follows constant velocity motion in $\Real^{2,2}$ with its velocity $\sqrt{ 2 \lambda} =  \sqrt{\dot{W}^2} $.   
Then the geodesic distance $\sigma$ between the affine parameter $\tau_1$ and $\tau_2$  is given by 
\be
\sigma = \sqrt{\dot{W}^2}(\tau_1-\tau_2) \,.
\ee

If we define a parameter $\xi$ as
\begin{align}
\xi^{-1}\equiv - W^A(\tau_1) W^B(\tau_2) \, \eta_{AB} 
&= \cosh\left[\sqrt{\dot{W}^2}(\tau_1-\tau_2)\right]  \nonumber \\
&= \cosh \left [ \sigma(X(\tau_1), X(\tau_2)) \right] \,,
\end{align} 
where we used \eqref{geosolu} and \eqref{cconstraint}, 
then, 
\be
\sigma(X(\tau_1), X(\tau_2))=\cosh^{-1}\left(\frac{1}{\xi}\right)=\log\left(\dfrac{1+\sqrt{1-\xi^2}}{\xi}\right) \,.
\ee

\subsection{Matrix representation of \AdSt and its isometry} 
We will construct wormhole geometry by quotient (dividing our $t=0$ slice).   
For that purpose, it is more convenient to express the embedding formalism of \AdSt expressed in terms of the $2 \times 2$ matrices $\hat{M}$;   
\be
\label{AdSmatrix}
\hat{M} :=  
 \left(
    \begin{array}{cc}
      U+X & Y + V \\
      Y - V & U-X \\
    \end{array} 
  \right) \,, \,\, d\hat{M} =  
 \left(
    \begin{array}{cc}
      dU+ dX & dY + dV \\
      dY - dV & dU - dX \\
    \end{array} 
  \right) \,,
\ee
Then the $AdS_3$ is written simply as 
\begin{align}
ds^2 &= - \det d\hat{M} \,, \quad {\mbox{where}} \quad
 \det \hat{M}  = 1 \,.
\end{align}

 \underline{{\bf Isometry}}

This representation of \AdSt has manifest symmetry under the following transformation; 
\be
\label{generictr}
 \left(
    \begin{array}{cc}
      U+X & Y + V \\
      Y - V & U-X \\
    \end{array} 
  \right) \, 
\mapsto \, 
\gamma_1
 \left(
    \begin{array}{cc}
      U+X & Y + V \\
      Y - V & U-X \\
    \end{array} 
  \right)
\gamma_2^T\,,
\ee
where  $(\gamma_1,\gamma_2) \in SL(2,\Real)\times SL(2,\Real)$. 

We will consider the spacelike slice $t=0$, and divide that slice by the isometry, namely, we will construct wormholes as quotients of $V=0$ slice ($t=0$ slice) of the \AdSt. 
Furthermore, one can easily check that $V=0$ subspace is invariant under the transformation $\gamma_1 = \gamma_2 \equiv \gamma$. Therefore the isometry we will use is $\gamma  \in SL(2,\Real)$ transformation under which $V=0$ slice of the \AdSt transforms as 
\be
\label{eq:t=0tr}
 \left(
    \begin{array}{cc}
      U+X & Y  \\
      Y  & U-X \\
    \end{array} 
  \right) \, 
\mapsto \, 
\gamma
 \left(
    \begin{array}{cc}
      U+X & Y \\
      Y  & U-X \\
    \end{array} 
  \right)
\gamma^T\,.
\ee

To see how this transforms the hyperbolic space coordinate $(x, z)$, we set
\begin{equation}
\gamma = \left(
    \begin{array}{cc}
      a & b \\
      c  & d \\
    \end{array} 
  \right),
\end{equation}
with $ad - bc = 1$. Then using the relationship \eqref{tslicePoincare}, 
one can confirm by explicit calculation that the transformation of $(U, X, Y)$ under \eqref{eq:t=0tr} is exactly the following 
%
M\"{o}bius transformation of $Z \equiv x + i z$,   
\begin{align}
Z \to Z'  &= x' + i z'  = \frac{a Z + b }{c Z + d} 
= \f{ \{ (a x + b) (c x+ d ) + ac z^2 \}+ i z  }{\left( c x+ d \right)^2+  (c z)^2}\,.
\end{align}

In summary, the isometry \eqref{eq:t=0tr} transforms the Poincar\'e coordinate as following $t=0$ slice M\"{o}bius transformation 
\be
\label{Mobiustr}
\Bigl( 
t = 0 \,, 
Z = x + i z
\Bigr) 
\to 
\left( 
t'=0 \,, 
Z'   = \frac{a Z + b}{c Z + d} 
 \right)  \,.
\ee
An important point is that an $SL(2,\Real)$ transformation maps a circle to another circle, which we will review in the next subsection.

In terms of this matrix representation for two points $p$ and $q$ as $P$ and $Q$, where $P$ and $Q$ are matrix representation of point $p$, $q$ following \eqref{Poincarecoord} and \eqref{AdSmatrix},  
its length $\sigma(p, q)$ is given by the formula 
\be 
\cosh[\sigma (p,q)] = \f12 \Tr \left( P^{-1} \cdot Q \right) = \f12 \Tr \left( P \cdot Q^{-1} \right) \,.
\ee
This formula is manifestly invariance under the isometry \eqref{generictr}.

\subsection{Elements of $SL(2,\Real)$ and its transformations} 
\label{AppA4}
A M\"{o}bius transformation \eqref{Mobiustr} can be decomposed into three basic transformation elements: 1) dilatation, 2) translation,  and 3)  special conformal transformation.
\begin{enumerate}
\item Dilatation:   this corresponds to the M\"{o}bius transformation $g =  g_{D}(a)$ with 
  \be 
  g_{D}(a)=\left(
    \begin{array}{cc}
      a & 0 \\
      0 &  a^{-1} \\
    \end{array}
  \right),
  \ee
Under this, the hyperbolic space coordinate transforms as 
  \be
  Z \to a^2 Z   \,.  
  \ee
\item Translation: this corresponds to the M\"{o}bius transformation $g =  g_{T}(b)$ with  
\be
   g_{T}(b)=\left(
    \begin{array}{cc}
     1& b \\
      0 & 1 \\
    \end{array} 
  \right) \,. 
  \ee
Under this, the hyperbolic space coordinate transforms as 
  \be
  Z \to Z + b  \,.
  \ee
 \item Special conformal transformation:  this corresponds to the M\"{o}bius transformation $g =  g_{SC}(c)$ with 
 \be
   g_{SC}(c)=\left(
    \begin{array}{cc}
     1& 0 \\
      c &  1\\
    \end{array}
  \right)
  \ee
  Under this, the hyperbolic space coordinate transforms as 
  \be
  Z \to \frac{Z}{c Z + 1}   \,.
  \ee
Note that the spacial conformal transformation acts like 
\be 
x \rightarrow \f{x+c^{2}(x^{2}+z^{2})}{1+2cx+c^{2}(x^{2}+z^{2})}, \quad z \rightarrow  \f{z}{1+2 c x+c^{2}(x^{2}+z^{2})}.
\ee
\item  Inversion: 
In addition, it is also useful to define an inversion $I({R})$
\be 
I(R) \equiv g_{T} (R)\; g_{SC}\left(-\f{1}{R} \right) \; g_{T} (R)= \left(
    \begin{array}{cc}
     0& R \\
      -\f{1}{R} &  0\\
    \end{array}
  \right) 
\ee
Notice that the action of the inversion $I({R})$  is 
\be 
Z = x+iz \rightarrow -\f{R^{2}}{Z}  = R^{2} \left[-\f{x}{x^{2}+z^{2}} + i \f{z}{x^{2}+z^{2}} \right] \,, 
\ee
which maps the circle  $x^{2} +z^{2} =R^{2}$ to itself, but flips its orientation
$x \rightarrow -x$. 
This inversion maps the exterior of the circle $x^{2} +z^{2} >R^{2}$ to the interior  
$x^{2} +z^{2} <R^{2}$ . 
\end{enumerate}

An $SL(2,\Real)$ transformation maps a circle to another circle.  Since in \AdSt, spatial geodesic is semicircle, this means that $SL(2,\Real)$ transforms geodesic to another geodesic.  
Let $C_1, C_2$ be two semicircles on upper half plane, 
\be 
C_1: (x-c_{1})^{2} +z^{2} =R^{2}_{1} \,, \quad C_2: =(x-c_{2})^{2} +z^{2} =R^{2}_{2} \,. 
\label{eq:2circles}
\ee
Furthermore, suppose that two semicircles $C_1$ and $C_2$ have opposite orientations. 
Then the element of $SL(2,\Real)$
which maps the circle $C_1$ to $C_2$ is given by 
\begin{align} 
g&= g_{T}(c_{2}) \;g_{D}(a_{12})\; I({R_{1}})\; g_{T}^{-1}(c_{1}) \,, \quad \mbox{where} \quad a_{12} = \sqrt{\f{R_{2}}{R_{1}}}  \,.   
\label{eq:total}
\end{align}  
Accordingly, the total action $g$ \eqref{eq:total} maps interior of $C_1$ to exterior of $C_2$ and vice versa. An explicit expression for $g$ is 
\begin{align}
g = \left(
\begin{array}{cc}
 -\frac{c_2}{\sqrt{R_1 R_2}} & 
 \frac{c_1 c_2 + R_1 R_2}{\sqrt{R_1 R_2}}  \\
 -\frac{1}{\sqrt{R_1 R_2}} &
 \frac{c_1}{\sqrt{R_1 R_2}}\\
\end{array}
\right) \,.
\end{align} 
For a more detailed treatment, see also \cite{Caceres:2019giy}.

\subsection{Length of closed geodesics from group elements}\label{App5}

\begin{figure}[t]
 \begin{center}
  \resizebox{120mm}{!}{\includegraphics{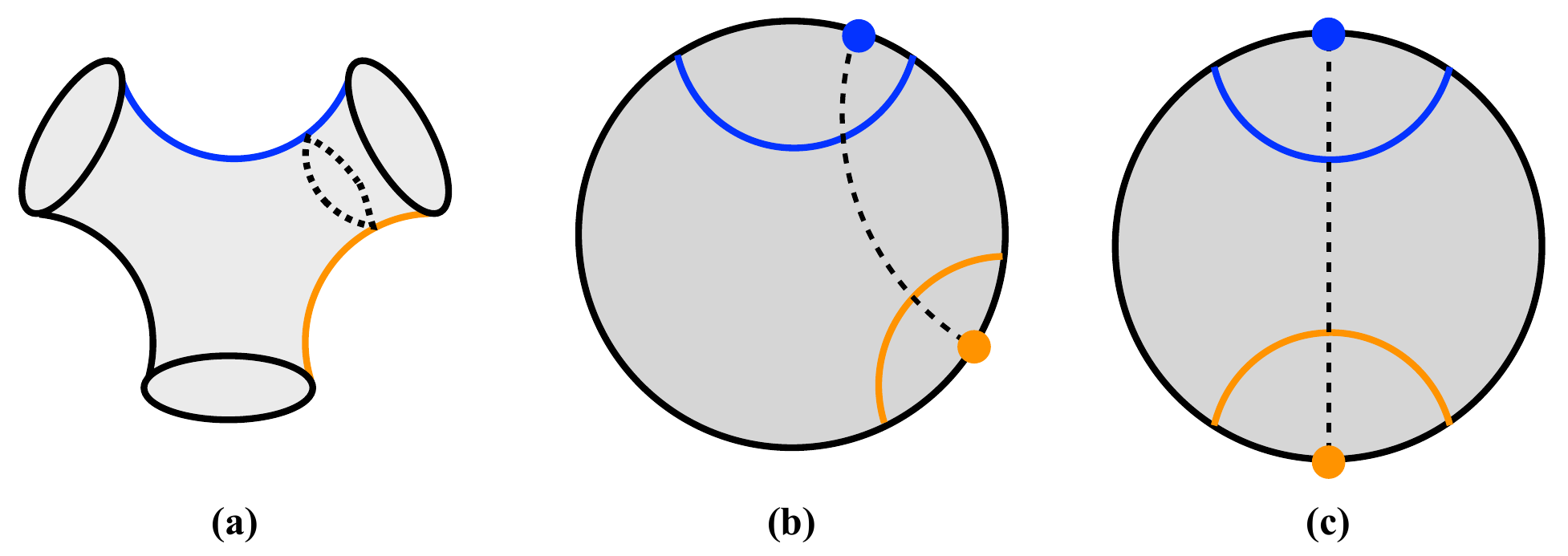}}
 \end{center}
 \caption{(a) We would like to compute the length of the closed geodesic (dotted line) in the wormhole geometry. (b) We can relate this calculation to the one in the Poincare disk. There are two fixed points of our geodesics and these correspond to the fixed points of a group element $\gamma$. (c) We can diagonalize $\gamma$ or equivalently we can move these two points to antipodal points. Then, our calculation of closed geodesics reduce to the calculation of the $\mathrm{Tr}\gamma$.}
 \label{fig:cgeo}
\end{figure}

In the main body, we have seen the length of closed geodesics in the eternal BTZ can be mapped to the calculation of trace of identification group (see equation \eqref{eq:Lbtz}). Here we explain the similar formula \eqref{eq:formula} for more general closed geodesics, characterized by an identification group $\gamma$. 

In the case of eternal BTZ, we computed the length of geodesics from a straight line anchored on the bulk points as \eqref{eq:line}. Then, we argued this calculation is equivalent to calculate the trace of $\gamma_1$ as \eqref{eq:Lbtz}. This group element $\gamma_1$ specifies two boundary points as follows. In the hyperbolic space $\mathbb{H}^2$, we have two fixed point of this dilatation $\gamma_1$. To make illustration simpler, let us map our upper half-plane to the Poincare disk. Then, these two fixed points are antipodal points on the boundary $S^1$. In particular, the straight line of our interest can be uniquely extended to these two boundary points. One may see (c) of Figure \ref{fig:cgeo} as a reference. 

Having this fact in mind, let us start from a given closed geodesics constructed from the Poincare disk and its identification. (One may refer to (a) and (b) of Figure \ref{fig:cgeo}.) Then, we can extend our geodesics anchored on the bulk points to the one anchored on two boundary points. Notice that this specification of boundary points should determine the corresponding group element, say $\gamma$, as we can always diagonalize it. Thanks to the isometry of AdS, we can map these boundary fixed points to antipodal ones as like the BTZ example. After all, our calculation reduces to the previous BTZ calculation, thus \eqref{eq:formula} holds.


\begin{thebibliography}{99}
\bibitem{Coleman:1988cy}
S.~R.~Coleman,
``Black Holes as Red Herrings: Topological Fluctuations and the Loss of Quantum Coherence,''
Nucl. Phys. B \textbf{307}, 867-882 (1988)
doi:10.1016/0550-3213(88)90110-1

\bibitem{Hawking:1987mz}
S.~W.~Hawking,
``Quantum Coherence Down the Wormhole,''
Phys. Lett. B \textbf{195}, 337 (1987)
doi:10.1016/0370-2693(87)90028-1

\bibitem{Hawking:1988ae}
S.~W.~Hawking,
``Wormholes in Space-Time,''
Phys. Rev. D \textbf{37}, 904-910 (1988)
doi:10.1103/PhysRevD.37.904

\bibitem{Giddings:1987cg}
S.~B.~Giddings and A.~Strominger,
``Axion Induced Topology Change in Quantum Gravity and String Theory,''
Nucl. Phys. B \textbf{306}, 890-907 (1988)
doi:10.1016/0550-3213(88)90446-4

\bibitem{Giddings:1988cx}
S.~B.~Giddings and A.~Strominger,
``Loss of Incoherence and Determination of Coupling Constants in Quantum Gravity,''
Nucl. Phys. B \textbf{307}, 854-866 (1988)
doi:10.1016/0550-3213(88)90109-5

\bibitem{Giddings:1988wv}
S.~B.~Giddings and A.~Strominger,
``Baby Universes, Third Quantization and the Cosmological Constant,''
Nucl. Phys. B \textbf{321}, 481-508 (1989)
doi:10.1016/0550-3213(89)90353-2

\bibitem{Lavrelashvili:1987jg}
G.~V.~Lavrelashvili, V.~A.~Rubakov and P.~G.~Tinyakov,
``Disruption of Quantum Coherence upon a Change in Spatial Topology in Quantum Gravity,''
JETP Lett. \textbf{46}, 167-169 (1987)

\bibitem{Hebecker:2018ofv}
A.~Hebecker, T.~Mikhail and P.~Soler,
``Euclidean wormholes, baby universes, and their impact on particle physics and cosmology,''
Front. Astron. Space Sci. \textbf{5}, 35 (2018)
doi:10.3389/fspas.2018.00035
[arXiv:1807.00824 [hep-th]].


\bibitem{Maldacena:2004rf}
J.~M.~Maldacena and L.~Maoz,
``Wormholes in AdS,''
JHEP \textbf{02}, 053 (2004)
doi:10.1088/1126-6708/2004/02/053
[arXiv:hep-th/0401024 [hep-th]].

\bibitem{Saad:2019lba}
P.~Saad, S.~H.~Shenker and D.~Stanford,
``JT gravity as a matrix integral,''
[arXiv:1903.11115 [hep-th]].

\bibitem{Marolf:2020xie}
D.~Marolf and H.~Maxfield,
``Transcending the ensemble: baby universes, spacetime wormholes, and the order and disorder of black hole information,''
JHEP \textbf{08}, 044 (2020)
doi:10.1007/JHEP08(2020)044
[arXiv:2002.08950 [hep-th]].

\bibitem{ArkaniHamed:2007js}
N.~Arkani-Hamed, J.~Orgera and J.~Polchinski,
``Euclidean wormholes in string theory,''
JHEP \textbf{12}, 018 (2007)
doi:10.1088/1126-6708/2007/12/018
[arXiv:0705.2768 [hep-th]].

\bibitem{Balasubramanian:2020jhl}
V.~Balasubramanian, A.~Kar, S.~F.~Ross and T.~Ugajin,
``Spin structures and baby universes,''
JHEP \textbf{09}, 192 (2020)
doi:10.1007/JHEP09(2020)192
[arXiv:2007.04333 [hep-th]].

\bibitem{Gardiner:2020vjp}
J.~G.~Gardiner and S.~Megas,
``2d TQFT and baby universes,''
[arXiv:2011.06137 [hep-th]].


\bibitem{Gesteau:2020wrk}
E.~Gesteau and M.~J.~Kang,
``Holographic baby universes: an observable story,''
[arXiv:2006.14620 [hep-th]].

\bibitem{Belin:2020hea}
A.~Belin and J.~de Boer,
``Random Statistics of OPE Coefficients and Euclidean Wormholes,''
[arXiv:2006.05499 [hep-th]].

\bibitem{Maloney:2020nni}
A.~Maloney and E.~Witten,
``Averaging over Narain moduli space,''
JHEP \textbf{10}, 187 (2020)
doi:10.1007/JHEP10(2020)187
[arXiv:2006.04855 [hep-th]].

\bibitem{Afkhami-Jeddi:2020ezh}
N.~Afkhami-Jeddi, H.~Cohn, T.~Hartman and A.~Tajdini,
``Free partition functions and an averaged holographic duality,''
[arXiv:2006.04839 [hep-th]].

\bibitem{Cotler:2020ugk}
J.~Cotler and K.~Jensen,
``AdS$_3$ gravity and random CFT,''
[arXiv:2006.08648 [hep-th]].

\bibitem{Altland:2020ccq}
A.~Altland and J.~Sonner,
``Late time physics of holographic quantum chaos,''
[arXiv:2008.02271 [hep-th]].

\bibitem{Maldacena:2001kr}
J.~M.~Maldacena,
``Eternal black holes in anti-de Sitter,''
JHEP \textbf{04}, 021 (2003)
doi:10.1088/1126-6708/2003/04/021
[arXiv:hep-th/0106112 [hep-th]].

\bibitem{Israel:1976ur}
W.~Israel,
``Thermo field dynamics of black holes,''
Phys. Lett. A \textbf{57}, 107-110 (1976)
doi:10.1016/0375-9601(76)90178-X

\bibitem{Balasubramanian:1998de}
V.~Balasubramanian, P.~Kraus, A.~E.~Lawrence and S.~P.~Trivedi,
``Holographic probes of anti-de Sitter space-times,''
Phys. Rev. D \textbf{59}, 104021 (1999)
doi:10.1103/PhysRevD.59.104021
[arXiv:hep-th/9808017 [hep-th]].

\bibitem{Maldacena:2013xja}
J.~Maldacena and L.~Susskind,
``Cool horizons for entangled black holes,''
Fortsch. Phys. \textbf{61}, 781-811 (2013)
doi:10.1002/prop.201300020
[arXiv:1306.0533 [hep-th]].

\bibitem{VanRaamsdonk:2010pw}
M.~Van Raamsdonk,
``Building up spacetime with quantum entanglement,''
Gen. Rel. Grav. \textbf{42}, 2323-2329 (2010)
doi:10.1142/S0218271810018529
[arXiv:1005.3035 [hep-th]].




\bibitem{Verlinde:2020upt}
H.~Verlinde,
``ER = EPR revisited: On the Entropy of an Einstein-Rosen Bridge,''
[arXiv:2003.13117 [hep-th]].


\bibitem{Brill:1998pr}
D.~Brill,
``Black holes and wormholes in (2+1)-dimensions,''
Lect. Notes Phys. \textbf{537}, 143 (2000)
[arXiv:gr-qc/9904083 [gr-qc]].

\bibitem{Skenderis:2009ju}
K.~Skenderis and B.~C.~van Rees,
``Holography and wormholes in 2+1 dimensions,''
Commun. Math. Phys. \textbf{301}, 583-626 (2011)
doi:10.1007/s00220-010-1163-z
[arXiv:0912.2090 [hep-th]].

\bibitem{Aminneborg:1997pz}
S.~Aminneborg, I.~Bengtsson, D.~Brill, S.~Holst and P.~Peldan,
``Black holes and wormholes in (2+1)-dimensions,''
Class. Quant. Grav. \textbf{15}, 627-644 (1998)
doi:10.1088/0264-9381/15/3/013
[arXiv:gr-qc/9707036 [gr-qc]].

\bibitem{Balasubramanian:2014hda}
V.~Balasubramanian, P.~Hayden, A.~Maloney, D.~Marolf and S.~F.~Ross,
``Multiboundary Wormholes and Holographic Entanglement,''
Class. Quant. Grav. \textbf{31}, 185015 (2014)
doi:10.1088/0264-9381/31/18/185015
[arXiv:1406.2663 [hep-th]].

\bibitem{Maxfield:2014kra}
H.~Maxfield,
``Entanglement entropy in three dimensional gravity,''
JHEP \textbf{04}, 031 (2015)
doi:10.1007/JHEP04(2015)031
[arXiv:1412.0687 [hep-th]].

\bibitem{Zhang:2016evx}
J.~d.~Zhang and B.~Chen,
``Kinematic Space and Wormholes,''
JHEP \textbf{01}, 092 (2017)
doi:10.1007/JHEP01(2017)092
[arXiv:1610.07134 [hep-th]].

\bibitem{Balasubramanian:2020hfs}
V.~Balasubramanian, A.~Kar, O.~Parrikar, G.~S\'arosi and T.~Ugajin,
``Geometric secret sharing in a model of Hawking radiation,''
[arXiv:2003.05448 [hep-th]].

\bibitem{Li:2020ceg}
T.~Li, J.~Chu and Y.~Zhou,
``Reflected Entropy for an Evaporating Black Hole,''
[arXiv:2006.10846 [hep-th]].


\bibitem{Dai:2020ffw}
D.~C.~Dai, D.~Minic, D.~Stojkovic and C.~Fu,
``Testing the $\mathbf {ER=EPR}$ conjecture,''
Phys. Rev. D \textbf{102}, no.6, 066004 (2020)
doi:10.1103/PhysRevD.102.066004
[arXiv:2002.08178 [hep-th]].

\bibitem{Shenker:2014cwa}
S.~H.~Shenker and D.~Stanford,
``Stringy effects in scrambling,''
JHEP \textbf{05}, 132 (2015)
doi:10.1007/JHEP05(2015)132
[arXiv:1412.6087 [hep-th]].

\bibitem{Maldacena:2015waa}
J.~Maldacena, S.~H.~Shenker and D.~Stanford,
``A bound on chaos,''
JHEP \textbf{08}, 106 (2016)
doi:10.1007/JHEP08(2016)106
[arXiv:1503.01409 [hep-th]].

\bibitem{Hayden:2011ag}
P.~Hayden, M.~Headrick and A.~Maloney,
``Holographic Mutual Information is Monogamous,''
Phys. Rev. D \textbf{87}, no.4, 046003 (2013)
doi:10.1103/PhysRevD.87.046003
[arXiv:1107.2940 [hep-th]].


\bibitem{Takayanagi:2017knl}
T.~Takayanagi and K.~Umemoto,
``Entanglement of purification through holographic duality,''
Nature Phys. \textbf{14}, no.6, 573-577 (2018)
doi:10.1038/s41567-018-0075-2
[arXiv:1708.09393 [hep-th]].



\bibitem{Nguyen:2017yqw}
P.~Nguyen, T.~Devakul, M.~G.~Halbasch, M.~P.~Zaletel and B.~Swingle,
``Entanglement of purification: from spin chains to holography,''
JHEP \textbf{01}, 098 (2018)
doi:10.1007/JHEP01(2018)098
[arXiv:1709.07424 [hep-th]].
\bibitem{Kudler-Flam:2018qjo}
J.~Kudler-Flam and S.~Ryu,
``Entanglement negativity and minimal entanglement wedge cross sections in holographic theories,''
Phys. Rev. D \textbf{99}, no.10, 106014 (2019)
doi:10.1103/PhysRevD.99.106014
[arXiv:1808.00446 [hep-th]].
\bibitem{Tamaoka:2018ned}
K.~Tamaoka,
``Entanglement Wedge Cross Section from the Dual Density Matrix,''
Phys. Rev. Lett. \textbf{122}, no.14, 141601 (2019)
doi:10.1103/PhysRevLett.122.141601
[arXiv:1809.09109 [hep-th]].

\bibitem{Dutta:2019gen}
S.~Dutta and T.~Faulkner,
``A canonical purification for the entanglement wedge cross-section,''
[arXiv:1905.00577 [hep-th]].

\bibitem{Vidal:2002zz}
G.~Vidal and R.~F.~Werner,
``Computable measure of entanglement,''
Phys. Rev. A \textbf{65}, 032314 (2002)
doi:10.1103/PhysRevA.65.032314
[arXiv:quant-ph/0102117 [quant-ph]].

\bibitem{Takayanagi:2018zqx}
T.~Takayanagi, T.~Ugajin and K.~Umemoto,
``Towards an Entanglement Measure for Mixed States in CFTs Based on Relative Entropy,''
JHEP \textbf{10}, 166 (2018)
doi:10.1007/JHEP10(2018)166
[arXiv:1807.09448 [hep-th]].

\bibitem{Hartman:2013mia}
T.~Hartman,
``Entanglement Entropy at Large Central Charge,''
[arXiv:1303.6955 [hep-th]].

\bibitem{Kusuki:2019rbk}
Y.~Kusuki and K.~Tamaoka,
``Dynamics of Entanglement Wedge Cross Section from Conformal Field Theories,''
[arXiv:1907.06646 [hep-th]].

\bibitem{Kusuki:2019evw}
Y.~Kusuki and K.~Tamaoka,
``Entanglement Wedge Cross Section from CFT: Dynamics of Local Operator Quench,''
JHEP \textbf{02}, 017 (2020)
doi:10.1007/JHEP02(2020)017
[arXiv:1909.06790 [hep-th]].

\bibitem{Asplund:2014coa}
C.~T.~Asplund, A.~Bernamonti, F.~Galli and T.~Hartman,
``Holographic Entanglement Entropy from 2d CFT: Heavy States and Local Quenches,''
JHEP \textbf{02}, 171 (2015)
doi:10.1007/JHEP02(2015)171
[arXiv:1410.1392 [hep-th]].

\bibitem{Srednicki}
M.~Srednicki, ``The Approach to Thermal Equilibrium in Quantized Chaotic Systems,'' J. Phys. A {\bf 32} (1999) 1163, [arXiv:cond-mat/9809360].

\bibitem{delCampo:2019qdx}
A.~Del Campo and T.~Takayanagi,
``Decoherence in Conformal Field Theory,''
JHEP \textbf{02}, 170 (2020)
doi:10.1007/JHEP02(2020)170
[arXiv:1911.07861 [hep-th]].


\bibitem{Sachdev:1992fk}
S.~Sachdev and J.~Ye,
``Gapless spin fluid ground state in a random, quantum Heisenberg magnet,''
Phys. Rev. Lett. \textbf{70}, 3339 (1993)
doi:10.1103/PhysRevLett.70.3339
[arXiv:cond-mat/9212030 [cond-mat]].

\bibitem{Kit}
A. Kitaev, A simple model of quantum holography, Talks at KITP, April 7, 2015 and May 27,
2015

\bibitem{Almheiri:2019jqq}
A.~Almheiri, A.~Milekhin and B.~Swingle,
``Universal Constraints on Energy Flow and SYK Thermalization,''
[arXiv:1912.04912 [hep-th]].



\bibitem{Piroli:2020dlx}
L.~Piroli, C.~S\"underhauf and X.~L.~Qi,
``A Random Unitary Circuit Model for Black Hole Evaporation,''
JHEP \textbf{04}, 063 (2020)
doi:10.1007/JHEP04(2020)063
[arXiv:2002.09236 [hep-th]].

\bibitem{Bengtsson1998}
I.~Bengtsson,
``ANTI-DE SITTER SPACE'',
http://3dhouse.se/ingemar/relteori/Kurs.pdf .

\bibitem{Caceres:2019giy}
E.~Caceres, A.~Kundu, A.~K.~Patra and S.~Shashi,
``A Killing Vector Treatment of Multiboundary Wormholes,''
JHEP \textbf{02}, 149 (2020)
doi:10.1007/JHEP02(2020)149
[arXiv:1912.08793 [hep-th]].

\end{thebibliography}
\end{document}